\documentclass[a4paper,11pt]{article}
\pdfoutput=1
\usepackage{jheppub}

\usepackage{graphicx}
\usepackage{dcolumn}
\usepackage{multirow}
\usepackage{bm}
\usepackage[T1]{fontenc}
\usepackage{feynmp-auto}
\usepackage{tabularx}
\usepackage{xcolor}
\usepackage{hyperref}
\usepackage{enumitem}

\usepackage[caption=false]{subfig}

\begin{document}
\title{Searching for Heavy Neutral Leptons at A Future Muon
Collider}
\author[a]{Tsz Hong Kwok,}
\author[b]{Lingfeng Li,}%
\author[a,c]{Tao Liu}
\author[c]{and Ariel Rock}
\affiliation[a]{Department of Physics, The Hong Kong University of Science and Technology, Clear Water Bay, Kowloon, Hong Kong S.A.R., PRC}
\affiliation[b]{Department of Physics, Brown University, Providence, RI, 02912, USA}
\affiliation[c]{Institute for Advanced Study, The Hong Kong University of Science and Technology, Clear Water Bay, Kowloon, Hong Kong S.A.R., PRC}
\emailAdd{thkwokae@connect.ust.hk}
\emailAdd{lingfeng\_li@brown.edu}
\emailAdd{taoliu@ust.hk}
\emailAdd{iasarock@ust.hk}
\date{\today}
\abstract{As the planning stages for a high energy muon collider enter a more concrete era, an important question arises as to what new physics could be uncovered. A TeV-scale muon collider is also a vector boson fusion (VBF) factory with a very clean background, and as such it is a promising environment to look for new physics that couples to the electroweak (EW) sector. In this paper, we explore the ability of a future TeV-scale muon collider to search for Majorana and Dirac Heavy Neutral Leptons (HNLs) produced via EW bosons. Employing a model-independent, conservative approach, we present an estimation of the production and decay rate of HNLs over a mass range between 200~GeV and 9.5~TeV in two benchmark collider proposals with $\sqrt{s}=3,\,10$~TeV, as well as an estimation of the dominant Standard Model (SM) background. We find that exclusion limits for the mixing between the HNLs and SM neutrinos can be as low as $\mathcal{O}(10^{-6})$. Additionally, we demonstrate that a TeV-scale muon collider allows for the ability to discriminate between Majorana and Dirac type HNLs for a large range of mixing values.}

\maketitle

\tableofcontents

\section{\label{sec:Intro}Introduction}
It is now well-known that active neutrinos oscillate and that at least two generations have very small but nonzero mass~\cite{homestake,kamland,dayabay,reno,doublechooz,snolab,ParticleDataGroup:2022pth}. As the Standard Model (SM) predicts that neutrinos have exactly zero mass, this is evidence of Beyond the Standard Model (BSM) physics. The question then arises as to what could be the origin of neutrino masses and mixings. It might be the case that new physics at an energy scale much higher than that of the SM may give rise to neutrino masses and mixings. If that new physics is renormalizable and allowed to violate lepton number, the lowest order effective operator at the scale of the SM is the $d=5$ dimension Weinberg operator~\cite{Weinberg:1979sa}. There are many possible ultraviolet (UV) theories that may result in this effective operator, but one such class of theories is called the Type I Seesaw mechanism~\cite{Gell-Mann:1979vob,Minkowski:1977sc,Yanagida:1979as,Mohapatra:1979ia}, in which the smallness of masses of the SM neutrinos is enforced by the presence of right-handed Heavy Neutral Leptons (HNLs) at a higher scale. 

To date, there have been numerous limits set on the coupling $|V_\ell|^2$ between HNLs and the SM. Prompt trilepton searches of HNL masses less than 60~GeV have been studied in hadron colliders, with sensitivity down to $\mathcal{O}(10^{-5})$~\cite{CMS:2018iaf,ATLAS:2019kpx}. Displaced searches in hadron colliders have been shown to have sensitivity reaching almost $\mathcal{O}(10^{-7})$ for HNL masses below 20~GeV~\cite{CMS:2022fut,ATLAS:2022atq}. Limits from LHCb also indicate sensitivity down to $\mathcal{O}(10^{-4})$ for masses below 50~GeV~\cite{LHCb:2020wxx}. Other future experiments promise even stricter limits, such as SHiP, with limits down to $\mathcal{O}(10^{-9})$ for masses below 5~GeV~\cite{SHiP:2018xqw}, or other potential hadron colliders~\cite{Pascoli:2018heg,Cottin:2021lzz}, with limits at $\mathcal{O}(10^{-7})$ for similar mass ranges. In addition, future lepton colliders show promise for higher mass HNLs, possibly down to $\mathcal{O}(10^{-5}-10^{-6})$ for masses between 200-2000~GeV~\cite{Mekala:2022cmm}. 

A future high-energy muon collider could serve as both an intensity and energy frontier, providing various opportunities for both direct and indirect searches of new physics. As compared to an electron-positron collider, a muon collider would have higher energy and luminosity, and reduced background~\cite{Skrinsky:1981ht,Neuffer:1983xya,Neuffer:1986dg,Barger:1995hr,Barger:1996jm,Ankenbrandt:1999cta}. As synchrotron radiation scales as $m^{-4},$ a muon collider would have an energy loss rate reduced by a  factor of $(m_e/m_\mu)^4\approx (207)^{-4}$ as compared to an electron-positron collider. This, in combination with the property that muon colliders would have negligible beamstrahlung~\cite{Chen:1996uk,Palmer:1996gs}, makes TeV-scale beam energies more feasible. Currently, estimates put the expected luminosity of a TeV-scale muon collider at order $1$~ab$^{-1}$~\cite{Aime:2022flm,Black:2022cth}. 

As compared to a proton collider, colliding muons would allow for a much more efficient probe of high energy processes, as a large portion of the beam energy would be carried by the muon itself, whereas a hadronic collider only admits high energy collisions at the tail of the proton's parton distribution function (PDF) --- which is highly suppressed. In fact, the efficiency gap is so pronounced that for some $2\to2$ processes, a muon collider with a given $\sqrt{s_\mu}$ has the same reach as a proton collider of $\sqrt{s_p}\sim 5-20\times\sqrt{s_\mu}$~\cite{Delahaye:2019omf,Costantini:2020stv,Aime:2022flm}. Furthermore, a muon collider is a perfect laboratory for studying new electroweak (EW) physics, such as in searches for HNLs. For a TeV-scale muon collider, roughly $5\%$ of the beam's energy is carried by EW bosons~\cite{Han:2020uid}, whereas a proton beam carries less than $1\%$~\cite{Bertone:2017bme}.

It is important to note, however, the various technical challenges facing the development of a muon collider. Unlike the beams of electron-positron or proton-proton colliders, muons are unstable, with a lifetime of around 2~$\mu$s. As such, it remains a formidable task to produce and store low emmittance muon beams. In recent years there have been rapid developments in overcoming these difficulties~\cite{Palmer:2014nza,MuonCollider:2022nsa,Schulte:2022cbw}. In the United States, the US Muon Accelerator Program (MAP) is investigating a potential muon source wherein a proton beam would collide with a high-Z fixed target to produce secondary muons as pion decay products~\cite{Delahaye:2013jla,Delahaye:2014vvd,Ryne:2015xua,Long:2020wfp}. However, the kinematics of this process admit final state muons with high emittance, and therefore significant cooling is needed. A possible cooling technology has been demonstrated by the Muon Ionization Cooling Experiment (MICE) at the Rutherford Appleton Laboratory in the United Kingdom~\cite{Mohayai:2018rxn,Blackmore:2018mfr,MICE:2019jkl}. An interesting approach to combine these two steps has been put forward in the Low Emmittance Muon Accelerator (LEMMA) proposal, which would produce muon pairs at threshold via a 45 GeV positron beam interacting with a fixed-target electron~\cite{Antonelli:2015nla}. These muons would have a long lifetime and small emittance. However, LEMMA is unable to achieve a high enough luminosity to be feasible~\cite{Black:2022cth}.

Nevertheless, it is clear that the new physics reach of a future muon collider is substantial, and the advantages over conventional lepton colliders have resulted in a multitude of phenomenological studies, such as searches for axion-like particles~\cite{Bao:2022onq,Han:2022mzp,Inan:2022rcr}, explorations of Higgs physics~\cite{Chiesa:2020awd,Han:2020pif,Chiesa:2021qpr,Cepeda:2021rql,Chen:2021pqi,Forslund:2022xjq,Buonincontri:2022ylv}, precision EW studies~\cite{DiLuzio:2018jwd,Buttazzo:2020uzc,Spor:2022mxl,Chen:2022msz,Chen:2022yiu}, flavor studies~\cite{Huang:2021biu,Asadi:2021gah,Haghighat:2021djz,Bandyopadhyay:2021pld,Das:2022mmh}, investigations of the muon $g-2~$\cite{Yin:2020afe}, and searches for dark matter~\cite{Han:2020uak,Capdevilla:2021fmj,Medina:2021ram,Black:2022qlg,Liu:2022byu}, among others~\cite{Costantini:2020stv,AlAli:2021let,Aime:2022flm,Homiller:2022iax}. Nevertheless, unlike for electron-positron colliders~\cite{Mekala:2022cmm}, the phenomenology of HNLs in a muon collider has not been studied in the literature. In this paper we present an analysis of the sensitivity and reach of a future muon collider in looking for HNLs.

This paper is structured as follows. In section~\ref{sec:Theory}, we introduce the the model we use that describes the interactions between the HNLs and the SM, and we describe the signal production channel of interest in the work.  In section~\ref{sec:Simulation}, we describe the framework we use to simulate the production and detection of signal and background events. In section~\ref{sec:Analysis}, we detail the reconstruction and analysis. Section~\ref{sec:Results} presents the predicted exclusion bounds on the coupling strengths of the HNLs as well as an estimation of the sensitivity in discerning between Majorana and Dirac HNLs. We conclude and consider future directions in section~\ref{sec:Conclusion}.

\section{\label{sec:Theory}Model}

From a collider physics perspective, an unmodified Type I Seesaw mechanism is excluded in the region of HNL mass and mixing parameter-space under consideration in this work. More specifically, for the TeV scale mass range of HNLs we focus on, the mixing $V_{N\ell}$ must be bounded below $10^{-6}$ in order to be compatible with current neutrino mass upper limits. Even for a future muon collider, such a mixing would be far too small to be probed. Thus, many collider studies focus on scenarios in which the HNLs couple to the SM via a new mediator with sizable couplings to the SM~\cite{Blanchet:2009bu,FileviezPerez:2009hdc,Anamiati:2016uxp,Deppisch:2015qwa,Chiang:2019ajm,Cottin:2021lzz,Liu:2021akf,Chauhan:2021xus}; in this case, the HNL production would rate become independent of the mixing parameters. This approach, however, is highly model-dependent. Another, more general, possible theory is a modification of the Type I Seesaw, the Inverse Seesaw mechanism~\cite{Mohapatra:1986bd,Mohapatra:1986aw,Nandi:1985uh}.
Via the introduction of small lepton number violating scale,  the Inverse Seesaw is able to accommodate the SM neutrino data for any values of mixing and HNL mass~\cite{Gonzalez-Garcia:1988okv,Bernabeu:1987gr}.

As the purpose of this work is to investigate the collider reach in observing electroweak production of HNLs in a model-independent fashion, we nevertheless remain agnostic as to the UV theory. Therefore, we employ an effective, phenomenological model --- the Universal FeynRules Output (UFO)~\cite{Degrande:2011ua,Alloul:2013bka} models \textit{HeavyN}, for Majorana~\cite{Alva:2014gxa,Degrande:2016aje,Atre:2009rg} and Dirac~\cite{Pascoli:2018heg} HNLs. In these implementations, the HNLs, $N_i$ for $i=1,\dots, 3$, are sterile in the SM but mix with the active neutrinos, namely,
\begin{equation}
\label{eq:mixing}
\nu_{\ell L} = \sum_{m=1}^3 U_{\ell m} \nu_{mL} + \sum_{m^\prime=1}^{3} V_{\ell m^\prime}N^c_{m^\prime L}~,
\end{equation}
where $m$, and $m'$ index mass eigenstates, and $\ell=e$, $\mu$, $\tau$ index gauge eigenstates. The matrix $U_{\ell m}$ is the usual Maki–Nakagawa–Sakata-Pontecorvo (MNSP) mixing matrix for SM neutrinos~\cite{Maki:1962mu,Pontecorvo:1957cp,Pontecorvo:1957qd}, and $V_{\ell m'}$ is a matrix parameterizing the mixing between the HNLs and the SM neutrinos.

\begin{figure}[tb]
\centering
\begin{fmfgraph*}(150,90)
\fmfleft{i}
\fmfright{o1,o2}
\fmf{boson,label=$W$,tension=1.5}{i,v1}
\fmf{plain,label=$\ell$,label.side=left}{v1,o2}
\fmf{plain,label=$N_i$}{o1,v1}
\end{fmfgraph*}
\qquad 
\begin{fmfgraph*}(150,90)
\fmfleft{i}
\fmfright{o1,o2}
\fmf{boson,label=$Z$,tension=1.5}{i,v1}
\fmf{plain,label=$\nu_\ell$,label.side=left}{v1,o2}
\fmf{plain,label=$N_i$}{o1,v1}
\end{fmfgraph*}

\caption{New interaction vertices between the HNLs and SM electroweak bosons in the simplified effective Lagrangian of equation~\ref{eq:lagrangian}.}
\label{fig:newverts}
\end{figure}
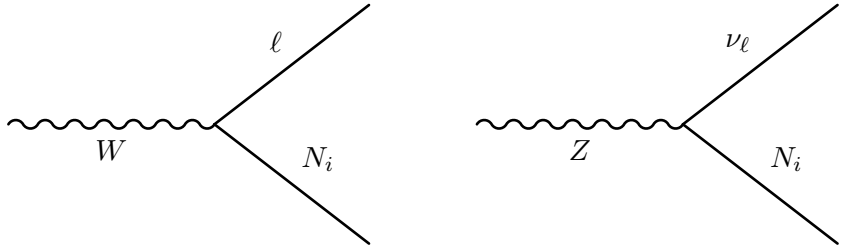

After electroweak symmetry breaking, the effective Lagrangian in the mass eigenstate basis that describes the interactions between the HNLs and the SM electroweak bosons is given by

\begin{align}
\label{eq:lagrangian}
-\mathcal{L}_{\text{int,EW}}=& \frac{g}{\sqrt{2}}W^{\mu+} \sum_{\ell=e}^\tau \bigg(\sum_{m=1}^3 U_{\ell m}^\ast \bar{\nu}_m\gamma^\mu P_L \ell +\sum_{m=1}^3 V_{\ell m}^\ast \bar{N}^c_m \gamma^\mu P_L \ell  \bigg)\\\nonumber
+&  \frac{g}{2\cos\theta_W}Z^{\mu} \sum_{\ell=e}^\tau \bigg(\sum_{m=1}^3 U_{\ell m}^\ast \bar{\nu}_m\gamma^\mu P_L \nu_\ell +\sum_{m=1}^3 V_{\ell m}^\ast \bar{N}^c_m \gamma^\mu P_L \nu_\ell  \bigg) +h.c.
\end{align}
This interaction Lagrangian introduces two new classes of vertices, given in figure~\ref{fig:newverts}. 

\begin{fmffile}{heavynverts}
\end{fmffile}

There are several production mechanisms for HNLs in a high-energy muon collider, the dominant of which depends on the mass of the HNL and its mixing with different generations. In the region of parameter space under consideration in this work, namely $\sqrt{s},m_{Ni}\gg m_W$, the dominant production process is the production of a HNL and SM neutrino pair. This $N_i \nu_\ell$ pair occurs via two production channels: either a $t$-channel exchange of a $W$ boson, or annihilation via a $Z$ boson. These channels (along with a decay $N\to q q \ell$) are shown in figure~\ref{fig:tchandiags} on the left- and right-hand sides, respectively. The $W$ exchange process benefits from a large $W$ boson electroweak (EW) parton distribution function (PDF)~\cite{Han:2020uid,Han:2021kes,Costantini:2020stv}. Conversely, the masses of the HNLs under consideration are much larger than $m_Z$, and as such annihilation processes via $Z$ decay are far off shell and highly suppressed. 

At $m_N\gg m_W$, we also have that the Goldstone Equivalence Theorem applies, and thus it then follows~\cite{Atre:2009rg,Pascoli:2018heg} that
\begin{equation}
\text{BR}(N\to W^+\ell^-)\approx 2 \text{BR}(N\to Z \nu_\ell) \approx 2\text{BR}(N\to h \nu_\ell).
\end{equation}
Note that this relation holds for both Majorana and Dirac HNLs, despite the absolute widths of the Majorana HNLs being twice as large as for the Dirac case. As the HNLs preferentially decay via charged-current interactions, and that the $W$ boson decays largely hadronically~\cite{ParticleDataGroup:2022pth}, we pick our decay channel of investigation to be $N\to q q \ell.$ An additional benefit to this channel is that all decay products are visible.

\section{\label{sec:Simulation}Simulation Framework}
\subsection{\label{subsec:Event}Signal Event Generation}

In this work we use the Monte-Carlo event generator \textsc{Whizard 3}~\cite{Kilian:2007gr,Moretti:2001zz} to generate events at the parton level. The advantage of using \textsc{Whizard 3} for this study is that it is possible to include the structure function for initial state radiation (ISR) for muon beams. Additionally, \textsc{Whizard 3} allows for the use of the equivalent photon approximation (EPA) for the inclusion of photon-induced background events due to the collinear splitting of $\mu\to\mu\gamma$. Furthermore, the implementations of ISR and EPA in \textsc{Whizard 3} allow for the insertion of $p_T$ recoil of the hard scattering processes into the event record. The following perturbative parton shower and hadronization steps are done using \textsc{Pythia 8}~\cite{Bierlich:2022pfr}.

\begin{figure}[tb]
\centering
\begin{fmfgraph*}(150,75)
\fmfstraight
\fmfleftn{i}{2}
\fmfstraight
\fmfrightn{o}{4}

\fmf{plain}{i1,v1}
\fmfv{label=$\mu$}{i1}
\fmfv{label=$\mu$}{i2}
\fmf{plain}{v1,o1}
\fmfv{label=$\nu$}{o1}
\fmf{plain}{i2,v2}
\fmf{plain,tension=2,label=$N$}{v2,v3}
\fmf{plain,tension=2}{v3,o4}
\fmf{boson,label=$W$,label.side=left}{v1,v2}
\fmffreeze
\fmf{boson,label=$W$}{v3,v4}
\fmf{plain}{o3,v4}
\fmf{plain,tension=2}{v4,o2}
\fmfv{label=$q$}{o2}
\fmfv{label=$q$}{o3}
\fmfv{label=$\ell$}{o4}
\end{fmfgraph*}
\qquad\qquad
\begin{fmfgraph*}(150,75)
\fmfstraight
\fmfleftn{i}{2}
\fmfstraight
\fmfrightn{o}{4}
\fmf{plain}{i1,v1}
\fmf{plain}{i2,v1}
\fmfv{label=$\mu$}{i1}
\fmfv{label=$\mu$}{i2}

\fmfv{label=$\nu$}{o1}
\fmf{boson,label=$Z$,label.side=right,tension=2}{v1,v2}
\fmf{plain}{v2,o1}
\fmf{plain,tension=2,label=$N$,label.side=left,tension=2}{v2,v3}
\fmf{plain,tension=2}{v3,o4}

\fmffreeze
\fmf{boson,label=$W$}{v3,v4}
\fmf{plain}{o3,v4}
\fmf{plain}{v4,o2}
\fmfv{label=$q$}{o2}
\fmfv{label=$q$}{o3}
\fmfv{label=$\ell$}{o4}
\end{fmfgraph*}
\caption{\label{fig:tchandiags}Feynman diagrams for the process $\mu^+\mu^-\to N (qq \ell^\pm)\nu$. For the collider energies above $m_Z$, such as the scenarios under consideration in this paper, the dominant production channel is given by the left-hand diagram of $t$-channel $W$ exchange.}
\end{figure}
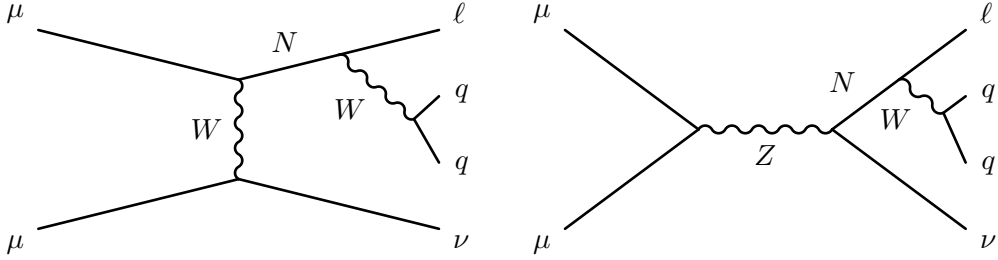

For the generation of signal events, we use the \textit{HeavyN} UFO files \textit{SM\_HeavyN\_NLO} and \textit{SM\_HeavyN\_Dirac\_NLO}, for Majorana and Dirac HNLs, respectively. We focus on the case in which only one HNL mixes with the SM. Furthermore, we assume that $|V_{1e}|=|V_{1\mu}|$ and $|V_{1\tau}|=0$. As we only consider one HNL, we use the notation $|V_{1 e}|=|V_{1\mu}|\equiv|V_{\ell}|$ with no ambiguity. We simulate the signal process by first generating the production of HNLs,  $\mu^+\mu^-\to N\nu,$ the Feynman diagrams of which are  shown in figure~\ref{fig:tchandiags}.

We then decay the HNL via $N\to qq\ell^\pm,$ where $q=u,\,d,\,c,\,s$ and $\ell^\pm=e^\pm,\,\mu^\pm.$  As we decay $N$ using the narrow-width approximation (NWA), we choose $|V_\ell|=0.002,$ as this enforces $\Gamma_{N}\ll m_N$.\footnote{This choice of $|V_\ell|$ is actually much smaller than required for the validity of the NWA, but, as noted in~\cite{Pascoli:2018heg}, our choice is advantageous in that larger mixings that still satisfy the NWA might allow for increased virtuality of the HNL, leading to large variations in the kinematic distributions of the HNL's decay products across events.} We are free to make this choice as $\sigma\left(\mu^+\mu^-\to N (qq \ell^\pm)\nu\right)/\left|V_{\ell}\right|^2$ is independent of $\left|V_\ell\right|$ for a given $m_N$. Furthermore, for the values of $\sqrt{s}$ and $m_N$ considered in this work, this choice of $|V_\ell|^2$ is still large enough to ensure prompt decays of the HNLs, as even in the most boosted, lowest width scenario under consideration (i.e $\sqrt{s}=10$~TeV and $m_N=200$~GeV), the decay length in the lab frame is of order $10^{-4}$ $\mu$m. This is significantly smaller than the anticipated spatial resolution of detectors under consideration in~\cite{Black:2022cth}. Note, however, that in non-trivial model extensions where the HNL production is governed by another operator or mediator, the HNL production rate may be independent of $V_\ell$. In such a case it is then possible to have $V_\ell \lesssim 10^{-6}$ while still producing an appreciable collider signature. In this case, it might be such that the decay length of HNL is macroscopic and the HNL becomes a long-lived particle~\cite{Deppisch:2015qwa,Deppisch:2019}. However, this approach deviates from the focus of this work, which is conservative and model-independent.

\begin{fmffile}{tchanprocess}
\end{fmffile}

In this work, we consider two of the muon collider benchmarks given in~\cite{Black:2022cth}: a scenario in which $\sqrt{s}=3$~TeV and $L=1$~ab$^{-1}$, and a scenario in which $\sqrt{s}=10$~TeV and $L=10$~ab$^{-1}$.  For analysis in a potential $3$~TeV (10~TeV) muon collider, we consider benchmark scenarios with masses $m_N$ between 200~GeV and 2900~GeV (200~GeV and 9500~GeV). 

\begin{figure}[tb]
    \centering
\includegraphics[width=\linewidth]{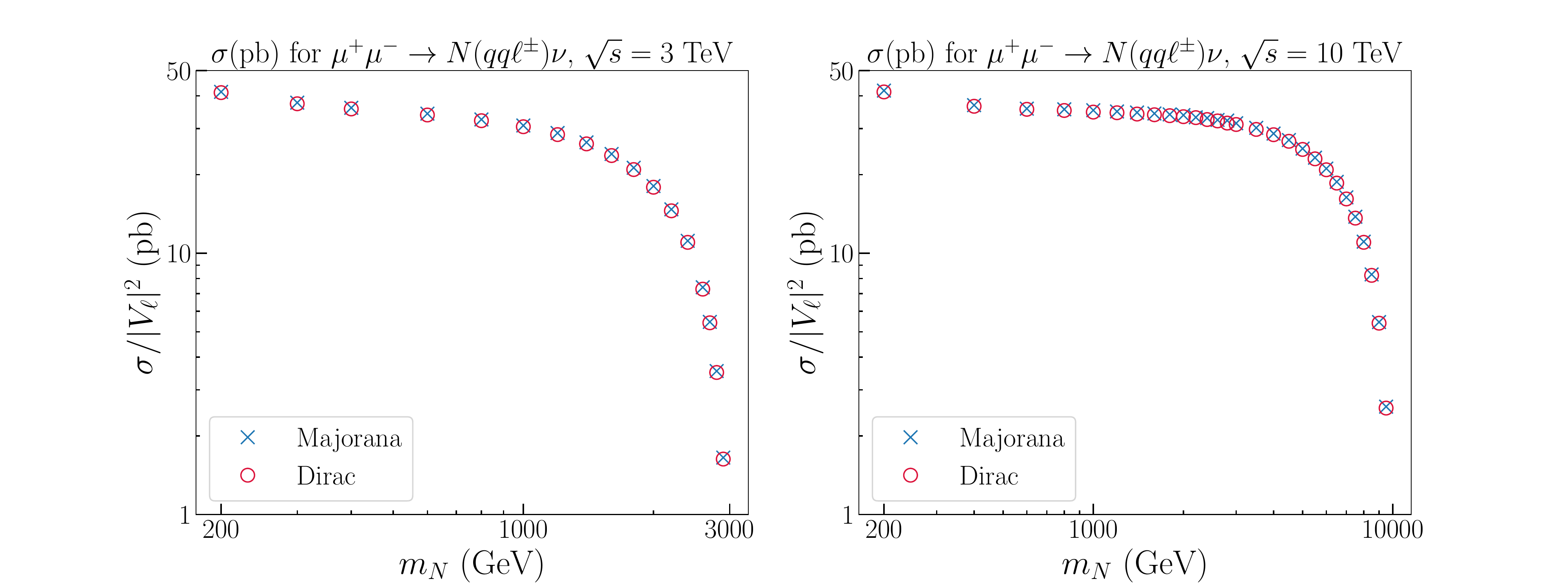}
    \caption{\label{fig:xsec_tchan} The total cross section divided by $|V_\ell|^2=|V_e|^2=|V_\mu|^2$ for the process $\mu^+\mu^-\to N(qq \ell^\pm)\nu$ as a function of $m_N$, shown at two benchmark future muon colliders with $\sqrt{s}=3,\,10$~TeV.}
\end{figure}

When generating signal events, we set all final state lepton and quark masses to zero. However, since we have ISR and ISR recoiled enabled, we set $m_\mu=0.105$~GeV for the beams, to be used as the mass for the ISR structure functions. At each mass benchmark, $\Gamma_N$ is recalculated and used to compute $\text{BR}(N\to q q \ell^\pm)$. The production cross section for each benchmark is then reweighted by $\text{BR}(N\to q q \ell^\pm)$ to give the total cross section for the process $\sigma(\mu^+\mu^-\to N (qq \ell^\pm)\nu)$. The resulting total process cross section divided by $|V_\ell|^2$ as a function of $m_N$ is shown in figure~\ref{fig:xsec_tchan}. Note that the cross section for $m_N \approx m_W$ is moderately enhanced by the branching of $N\to W^\pm\ell^\mp$ being favored over other decay channels at those masses, as noted in~\cite{Atre:2009rg}. 

\subsection{\label{subsec:bgevents}Background Event Generation}
As our final state topology of interest is two jets and a lepton, we consider SM background processes with two quarks and between one and three charged leptons in the final state. Specifically, we consider the following channels classified into two types according to the simulation methods used:
\begin{enumerate}[beginpenalty=10000]
\item[ME] \begin{enumerate}[beginpenalty=10000]
        \item $\mu^+\mu^-\to qq\ell\nu$,
        \item $\mu^+\mu^-\to qq\ell\ell$,
        \item $\mu^+\mu^-\to qq\ell\ell\nu\nu$,
        \item $\mu^+\mu^-\to qq\ell\ell\ell\nu$,
        \end{enumerate}
\item[EPA] \begin{enumerate}[resume]
        \item $\gamma\gamma\to qq\ell\nu$, 
        \item $\gamma\mu^\pm\to qq\ell$.
        \end{enumerate}
\end{enumerate}

The first four processes listed are computed using the full matrix element (ME). They include generator-level cuts such that the Monte-Carlo integration is convergent. In order to prevent underestimation of the SM background, the generator-level cuts are taken to be softer than those used in preselection. 

Furthermore, in order to preserve gauge-invariance, the simulations of the processes $\mu^+\mu^-\to qq\ell\ell\ell\nu$ and $\mu^+\mu^-\to qq\ell\ell$ include the processes $\gamma\gamma\to qq\ell\nu$ and $\gamma\mu^\pm\to qq\ell$ as subdiagrams. However, the cuts chosen on the full processes exclude the regions of low momentum transfer, $q_\ell$, where the contribution of partonic photon scattering dominates. Therefore, we include the last two processes, $\gamma\gamma\to qq\ell\nu$ and $\gamma\mu^\pm\to qq\ell$, where $\gamma$ refers to partonic photons split collinearly from the muon beam. These are simulated using the Equivalent Photon Approximation (EPA)~\cite{Budnev:1975poe}, based on the Weizs{\"a}cker-Williams approximation~\cite{vonWeizsacker:1934nji, Williams:1934ad}. We choose the lower cutoff $q_\ell$ for the full ME processes and the upper cutoff $q_\gamma$ for the EPA structure function to be complementary to avoid double counting of events.
We also include the effect of initial state radiation (ISR) on the beam(s) that do not split into EPA photons. The specific cuts used for the SM background processes are detailed in table~\ref{tab:cuts_gen_tchan}. The cross sections and estimated event yields for each background channel are given in table~\ref{tab:xsec_nevents}.

It is also worth commenting on the major contributors to each of the background channels. As expected, given that $\sqrt{s}$ is TeV scale, every background channel is dominated by contributions from vector boson fusion or scattering (VBF or VBS). More specifically, we find that the dominant processes in the channel $\mu^+\mu^-\to qq\ell\nu$ are from $WZ$ and $W\gamma$ to $W$ VBF, where the $W$ subsequently decays to a quark pair. The process $\mu^+\mu^-\to qq\ell\ell$ is dominated by $\gamma\gamma$ to $qq$ VBF via the exchange of a $t$-channel light quark. For $\mu^+\mu^-\to qq\ell\ell\nu\nu$, the major contribution is given by $W\gamma/WZ$ to $W\gamma/WZ$ VBS, where the neutral boson decays to a $qq$ pair, and the $W$ decays leptonically to $\ell\nu$. The channel $\mu^+\mu^-\to qq\ell\ell\ell\nu$ is dominated by $\gamma\gamma$ to $W W$ VBS, where one $W$ decays hadronically and the other leptonically. Additionally, the EPA process $\gamma\gamma\to qq\ell\nu$ is also dominated by the same $\gamma\gamma$ to $W W$ VBS, which is consistent with it being considered a subdiagram of the $\mu^+\mu^-\to qq\ell\ell\ell\nu$ channel. Likewise, the EPA process $\gamma\mu^\pm\to qq\ell$ is dominated by the same $\gamma\gamma$ to $qq$ VBS process that $\mu^+\mu^-\to qq\ell\ell$ is.
\begin{table*}
\centering

\begin{tabular}{|l|l|l|}
\hline
Process                     & Generator Level Cuts                                                                              & Method                    \\ \hline
$\mu^+\mu^-\to qq\ell\nu$   & \multirow{2}{*}{$M_{qq,\ell\ell} > 10$~GeV,\ $p_{T,\ell} > 4$~GeV, $|\eta_\ell|<8$, $q_\ell > 4$~GeV} & \multirow{2}{*}{ME + ISR} \\ \cline{1-1}
$\mu^+\mu^-\to qq\ell\ell$  &                                                                                                   &                           \\ \hline
$\mu^+\mu^-\to qq\ell\ell\nu\nu$   & \multirow{2}{*}{$M_{qq,\ell\ell} > 40$~GeV,\ $p_{T,\ell} > 4$~GeV, $|\eta_\ell|<8$, $q_\ell > 4$~GeV} & \multirow{2}{*}{ME + ISR} \\ \cline{1-1}
$\mu^+\mu^-\to qq\ell\ell\ell\nu$  &                                                                                                   &                           \\ \hline
$\gamma\gamma\to qq\ell\nu$ & $M_{qq} > 10$~GeV, $q_{\gamma} < 4$~GeV                                            & EPA     \\ \hline

$\gamma\mu^\pm\to qq\ell$       & $M_{qq} > 10$~GeV, $q_\gamma < 4$~GeV, $3 ^\circ < \theta_\ell < 177 ^\circ$                        &                  EPA + ISR         \\ \hline

\end{tabular}
\caption{\label{tab:cuts_gen_tchan} Summary of the cuts and simulation methods used to generate the SM background in \textsc{Whizard 3}. $q_\ell$ refers to the momentum-transfer between incoming and outgoing charged leptons, and $q_\gamma$ is the upper momentum-transfer for the EPA structure function. For the process $\gamma\mu^\pm\to qq\ell$, EPA is applied to one muon beam to produce the partonic photon, and ISR is applied to the other beam. For all processes, $p_T$ recoil on the hard scattering states due to emitted ISR (EPA) photons is applied to the generated events by setting ``\textsc{?Isr(Epa)\_Handler=True}'' and ``\textsc{?Isr(Epa)\_Handler\_Mode=Recoil}.'' }
\end{table*}

\subsection{\label{subsec:Detector}Detector Simulation}
After generating parton-level events with \textsc{Whizard 3}, and using \textsc{Pythia 8} to shower and hadronize, we use \textsc{Delphes 3}~\cite{deFavereau:2013fsa} to simulate the detector response. We use the default muon collider detector card that is included in the \textsc{Delphes 3} distribution. This card is a hybrid of the FCC-$hh$~\cite{Selvaggi:2717698} and CLIC~\cite{Roloff:2649439} cards. For this detector simulation, it requires isolated leptons to have $|\eta|<2.5$. 

It also includes a hypothetical forward muon spectrometer sensitive to muons with $2.5<|\eta|<8.0$ at $90\%$ efficiency for muons with $0.5$~GeV $<p_T<1.0$~GeV and at $95\%$ efficiency for muons with $2.5<|\eta|<8.0$ and $p_T>1.0$~GeV. This ability to detect forward muons is possibly advantageous in distinguishing signal from background in our analysis, as events that pass preselection with muons in the forward region are almost always due to the SM background. However, as the future of such a detector is uncertain, we therefore do not include its contribution in our analysis.

\section{\label{sec:Analysis}Analysis}

Following the detector simulation, we preselect events to keep those with signal-like topologies. Namely, in each event we require: 
\begin{itemize}
    \item \textbf{At least one isolated $\ell=e,\mu$ candidate.} Our definition of an isolated lepton is the same as the definition used by the built-in \textsc{Delphes} muon collider card. Firstly, it requires that the $\ell$ candidate have $p_T>0.5$~GeV. Secondly, inside a cone of $\Delta R=0.1$ around the $\ell$ candidate, the ratio of the scalar sum of $p_T$ of all particles in that cone to the $p_T$ of the $\ell$ candidate must be no more than 0.2. In case of having more than one $\ell$ candidate, the one with the highest $p_T$ will be used.
    \item \textbf{A $W$ boson jet system ($W_J$) candidate.}  We reconstruct the jet system in two possible ways, using the Valencia algorithm~\cite{Boronat:2014hva}. For most of the $m_N$ parameter space, $W$ bosons from the decays of the HNLs are highly boosted. Therefore, we choose to reconstruct the $W_J$ as one jet with a moderate radius parameter $R=1.2$ and both remaining parameters $\beta$ and $\gamma$ set to 1.0, denoted as $J_{\rm fat}$. 
    
    The second option is to reconstruct two narrower jets in a similar way, but with $R=0.2$, denoted as $j_1$ and $j_2$. If it is possible to reconstruct both classes of jet(s) in the event, we compare the invariant masses of $W_J=J_{\rm fat}$ with that of $W_J = j_1+j_2$. The jet reconstruction which gives an invariant mass closest to that of the $W$ boson is then accepted. If only one class of jet(s) is able to be reconstructed, that class will be used as $W_J$. 
    \item \textbf{Kinematic cuts on $\ell$ and $W_J$.} We require $\ell$ and $W_J$ to have $p_T>100$~GeV, and that the reconstructed jet system is on-shell, namely $|m_{W_J}-m_W|<5\Gamma_W$.
\end{itemize}

\begin{figure}[tb]
    \centering
   
    \includegraphics[width=\linewidth]{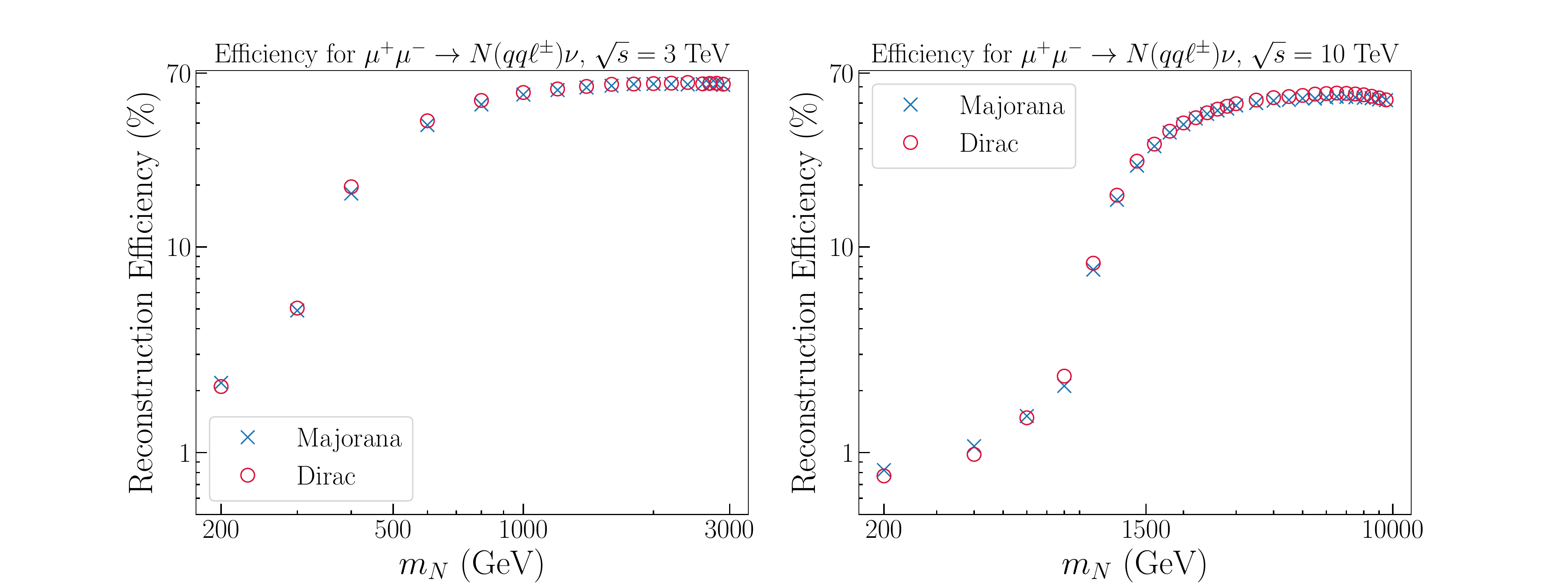}
    \caption{\label{fig:tchan_reco_eff}The preselection and reconstruction efficiency of the process $\mu^+\mu^-\to N (qq \ell^\pm)\nu$ as a function of $m_N$, shown at two benchmark future muon colliders with $\sqrt{s}=3,\,10$~TeV. As expected, the ability to detect and reconstruct events for HNLs with masses well below the collider energy is poor, as such HNLs are too boosted to have detectable decay products.}

\end{figure}

In figure~\ref{fig:tchan_reco_eff}, we show the reconstruction efficiency for Majorana and Dirac HNL signal channels. It is expected that the reconstruction efficiency is lessened for HNL masses well below $\sqrt{s},$ as such HNLs are mostly produced by the muon beam with a virtual $W$ boson with small $p_z$, as shown in the left panel of figure~\ref{fig:tchandiags}. The overall $p_z$ of the produced HNL becomes comparable to the beam energy, with the expected $p_T$ of its decay products limited by its small mass. Consequently the $\ell$ and $q\Bar{q}'$ final states are less likely to reach the central region with low $|\eta|$. The reconstruction efficiencies and event yields for the background channels are given in table~\ref{tab:xsec_nevents}. Note that the reconstruction efficiency of the $\mu^+\mu^-\to qq\ell\ell\ell\nu$ background channel is higher than that of the $\mu^+\mu^-\to qq\ell\nu$ channel. This is due to the fact that the major contributing process in the $\mu^+\mu^-\to qq\ell\ell\ell\nu$ background is $\mu^+\mu^-\to qq\ell\mu\mu\nu$, where the two outgoing muons are from the beam muons. The outgoing muons are therefore in the forward region, and will either be missed by the detector or detected with very small $p_T$. Such an event will still pass our reconstruction by identifying the remaining lepton (i.e. the one with highest $p_T$). Similarly, majority of the outgoing leptons in $\mu^+\mu^-\to qq\ell\nu$ channel are also beam muons. Therefore, this event will not pass preselection if such low $p_T$ forward muon is missed by the detector. Alternatively, one could also introduce a different reconstruction method which could strictly require that only one lepton is detected in the event in order to veto some of the $\mu^+\mu^-\to qq\ell\ell\ell\nu$ background. 

\begin{figure}[tb]
    \centering
    \includegraphics[width=0.65\linewidth]{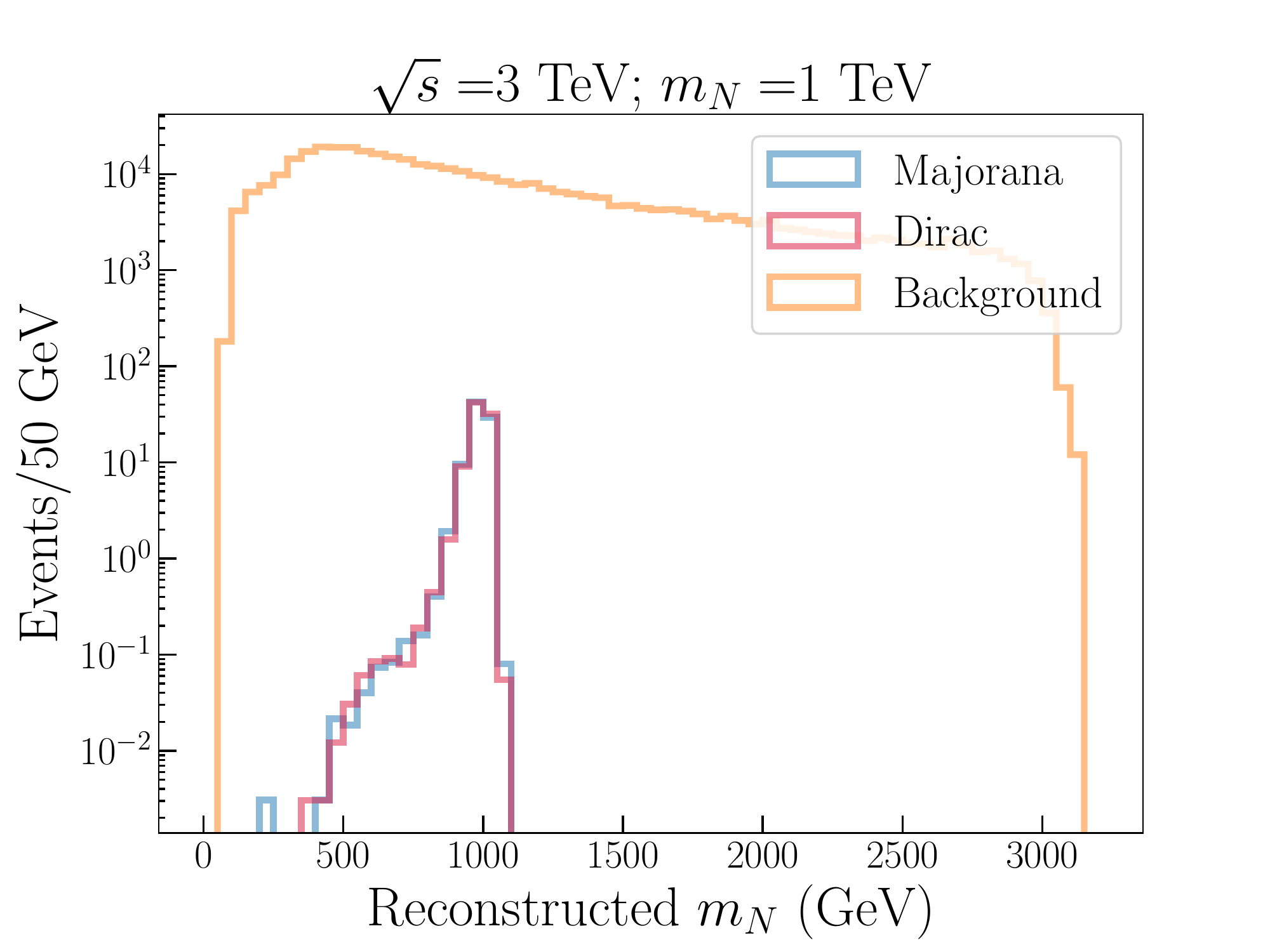}
    \caption{\label{fig:mn1plot}A histogram of reconstructed $m_N$ after preselection, for a benchmark scenario of $m_N=1$~TeV in a $\sqrt{s}=3$~TeV, $L=1$~ab$^{-1}$ muon collider. Blue corresponds to Majorana signal events, red for Dirac, and yellow for the combined SM background. Signal events are normalized to $|V_\ell|^2=5\times 10^{-6}.$ A sharp peak is visible at $m_N=1$~TeV with a tail to the left due missing $W$-jet components.}
 
\end{figure}

\begin{figure}[tb]
    \centering
    \includegraphics[width=\linewidth]{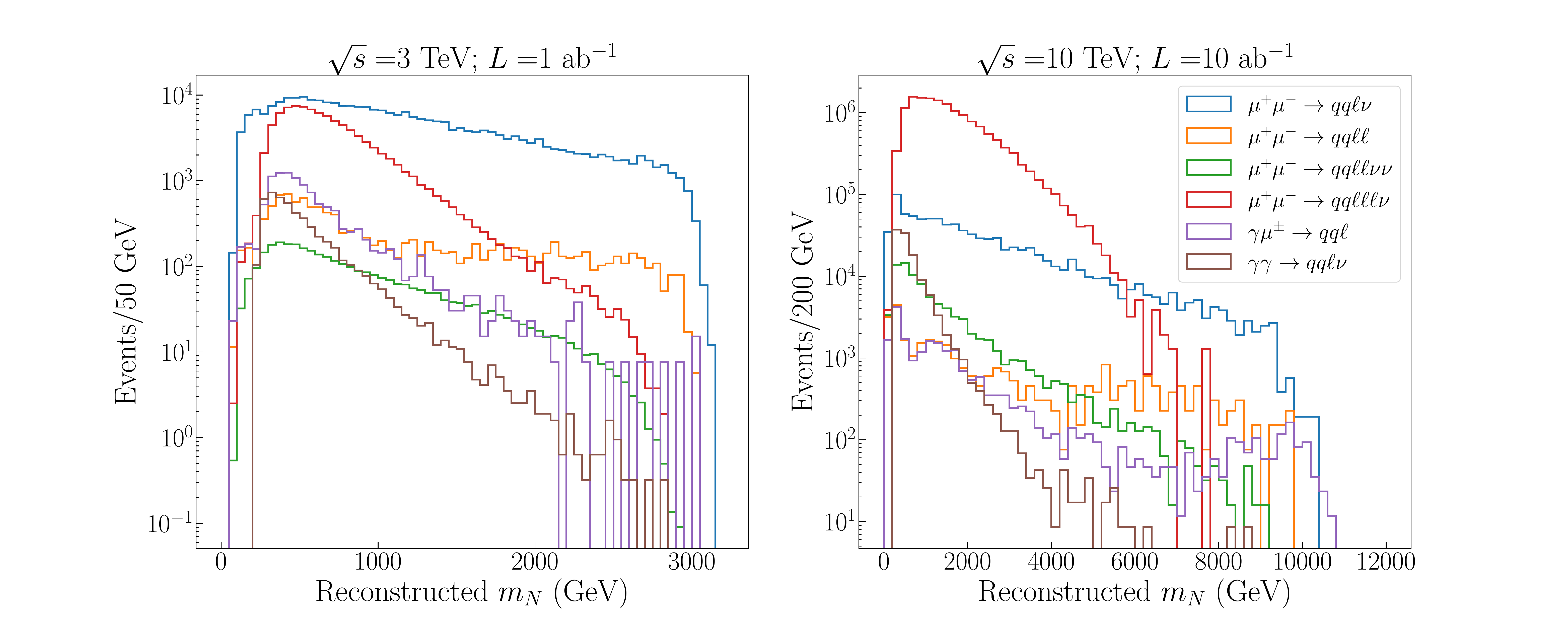}
    \caption{\label{fig:mn1plot_bg} The $m_N$ distribution for the individual channels that comprise the total SM background in both a $\sqrt{s}=3$~TeV (left) and $\sqrt{s}=10$~TeV (right) muon collider. For $\sqrt{s}=3$~TeV, we see that the dominant background channel is $\mu^+\mu^-\to qq \ell \nu$, whereas for $\sqrt{s}=10$~TeV the dominant background is due to the $\mu^+\mu^-\to qq\ell\ell\ell\nu$ channel for $m_N<6$~TeV, with the $\mu^+\mu^-\to qq \ell \nu$ channel becoming dominant for larger $m_N$.}
\end{figure}

After preselection, we then reconstruct the HNL by combining $W_J$ and the isolated lepton. A plot of the distribution of the invariant mass $m_N$ is show in figure~\ref{fig:mn1plot} for a benchmark scenario of $m_N=1$~TeV and $\sqrt{s}=3$~TeV. A peak in the signal events is visible at the expected value of $m_N=1$~TeV. Note that there is also a tail in the reconstructed mass distribution below the expected value. This is due to missing $W$-jet components that have either been missed by the detector reconstruction or not clustered by the jet algorithm. In both cases, the invariant mass reconstructed drops below the physical value. In figure~\ref{fig:mn1plot_bg}, we show the $m_N$ distribution for the individual channels that comprise the total SM background in both a $\sqrt{s}=3$~TeV and $\sqrt{s}=10$~TeV collider. On the left-hand side of figure~\ref{fig:mn1plot_bg}, for $\sqrt{s}=3$~TeV, we see that the dominant background channel is $\mu^+\mu^-\to qq \ell \nu$. For the right-hand side, with $\sqrt{s}=10$~TeV, the dominant background for $m_N<6$~TeV is due to the $\mu^+\mu^-\to qq\ell\ell\ell\nu$ channel, with the $\mu^+\mu^-\to qq \ell \nu$ becoming dominant for larger $m_N$.

To further separate signal-like and background-like events after preselection, we employ a Boosted Decision Tree (BDT) analysis using the Python implementation of \textit{XGBoost}~\cite{Chen:2016btl}. We train BDT to distinguish between three classes of event --- signal events due to the decay of a Majorana HNL, signal events due to the decay of a Dirac HNL, and signal events due to the SM background. The BDT training utilizes the following input features:
\begin{itemize}
    \item Lepton Information:
    \begin{itemize}
        \item The lepton's transverse momentum, $p_{T, \ell}$; pseudorapidity, $\eta_{\ell}$; and its energy, $E_\ell$.
        \item The charge of the lepton and its flavor.
    \end{itemize}
    \item $W$ Jet System Information:
    \begin{itemize}[beginpenalty=10000]
        \item The $W$ jet system's transverse momentum, $p_{T, W_J}$; pseudorapidity, $\eta_{W_J}$; and mass, $m_{W_J}$.
        \item The energies of the two narrow sub-jets $E_{j_1}$ and $E_{j_2}$, in the case that $W_J$ is able to be reconstructed as two jets.
    \end{itemize}
    \item Reconstructed HNL Information:
    \begin{itemize}
        \item The HNL's transverse and $z$-components of momenta, $p_{T, N}$ and $p_{z, N}$.
    \end{itemize}
    \item Geometric Information:
    \begin{itemize}
        \item  The angular distance and azimuthal angle difference between the lepton and $W$ jet system, $\Delta R(\ell, W_J)$ and $|\phi_\ell - \phi_{W_J}|$.
    \end{itemize}
\end{itemize}

\begin{table}[tb]
\centering
\begin{tabular}{l|lll|lll|}
\hline
\multicolumn{1}{|l|}{Collider COM Energy} & \multicolumn{3}{l|}{$\sqrt{s}=3$~TeV}                  & \multicolumn{3}{l|}{$\sqrt{s}=10$~TeV}                 \\ \hline
\multicolumn{1}{|l|}{Integrated Luminosity} & \multicolumn{3}{l|}{$L=1$~ab$^{-1}$}                  & \multicolumn{3}{l|}{$L=10$~ab$^{-1}$}                 \\ \hline\hline
\multicolumn{1}{|l|}{Process}                            & \multicolumn{1}{l|}{$\sigma$ (pb)} & \multicolumn{1}{l|}{$N_\text{events}$} & Eff. (\%) & \multicolumn{1}{l|}{$\sigma$ (pb)} & \multicolumn{1}{l|}{$N_\text{events}$} & Eff. (\%) \\ \hline
\multicolumn{1}{|l|}{$\mu^+\mu^-\to qq\ell\nu$}   & \multicolumn{1}{l|}{$6.025$}       &\multicolumn{1}{l|}{$263400$}          & 4.373 & \multicolumn{1}{l|}{$9.534$}       & \multicolumn{1}{l|}{$932800$}  & 0.9784        \\ \hline
\multicolumn{1}{|l|}{$\mu^+\mu^-\to qq\ell\ell$}  & \multicolumn{1}{l|}{$2.842$}       & \multicolumn{1}{l|}{$12160$}            & 0.4278 & \multicolumn{1}{l|}{$3.784$}       & \multicolumn{1}{l|}{$32090$}    & 0.0846       \\ \hline
\multicolumn{1}{|l|}{$\mu^+\mu^-\to qq\ell\ell\nu\nu$}   & \multicolumn{1}{l|}{$0.02255$}       &\multicolumn{1}{l|}{$3201$}          & 14.20 & \multicolumn{1}{l|}{$0.07968$}       & \multicolumn{1}{l|}{$85100$}  & 10.68        \\ \hline
\multicolumn{1}{|l|}{$\mu^+\mu^-\to qq\ell\ell\ell\nu$}  & \multicolumn{1}{l|}{$0.3133$}       & \multicolumn{1}{l|}{$90090$}            & 28.76 & \multicolumn{1}{l|}{$3.207$}       & \multicolumn{1}{l|}{$14950000$}    & 47.63      \\ \hline
\multicolumn{1}{|l|}{$\gamma\gamma\to qq\ell\nu$} & \multicolumn{1}{l|}{$0.1589$}      & \multicolumn{1}{l|}{$5068$}            & 3.190 & \multicolumn{1}{l|}{$0.4274$}      & \multicolumn{1}{l|}{$113600$}   & 2.658        \\ \hline
\multicolumn{1}{|l|}{$\gamma\mu^\pm\to qq\ell$}       & \multicolumn{1}{l|}{$3.811$}       & \multicolumn{1}{l|}{$11390$}   & 0.2986         & \multicolumn{1}{l|}{$0.5823$}      & \multicolumn{1}{l|}{$21360$}     &0.3668       \\ \hline

\end{tabular}
\caption{Cross sections, expected event yields and reconstruction efficiencies after the preselection for the SM background processes detailed in section~\ref{sec:Analysis}. Although the expected signal has only one final state charged lepton, we allow events with more than one charged lepton past preselection, as the Boosted Decision Tree analysis afterwards does an effective job of removing such events.}
\label{tab:xsec_nevents}
\end{table}

\begin{figure}[htb!]
    \centering
    \subfloat[]{\includegraphics[width=0.45\linewidth]{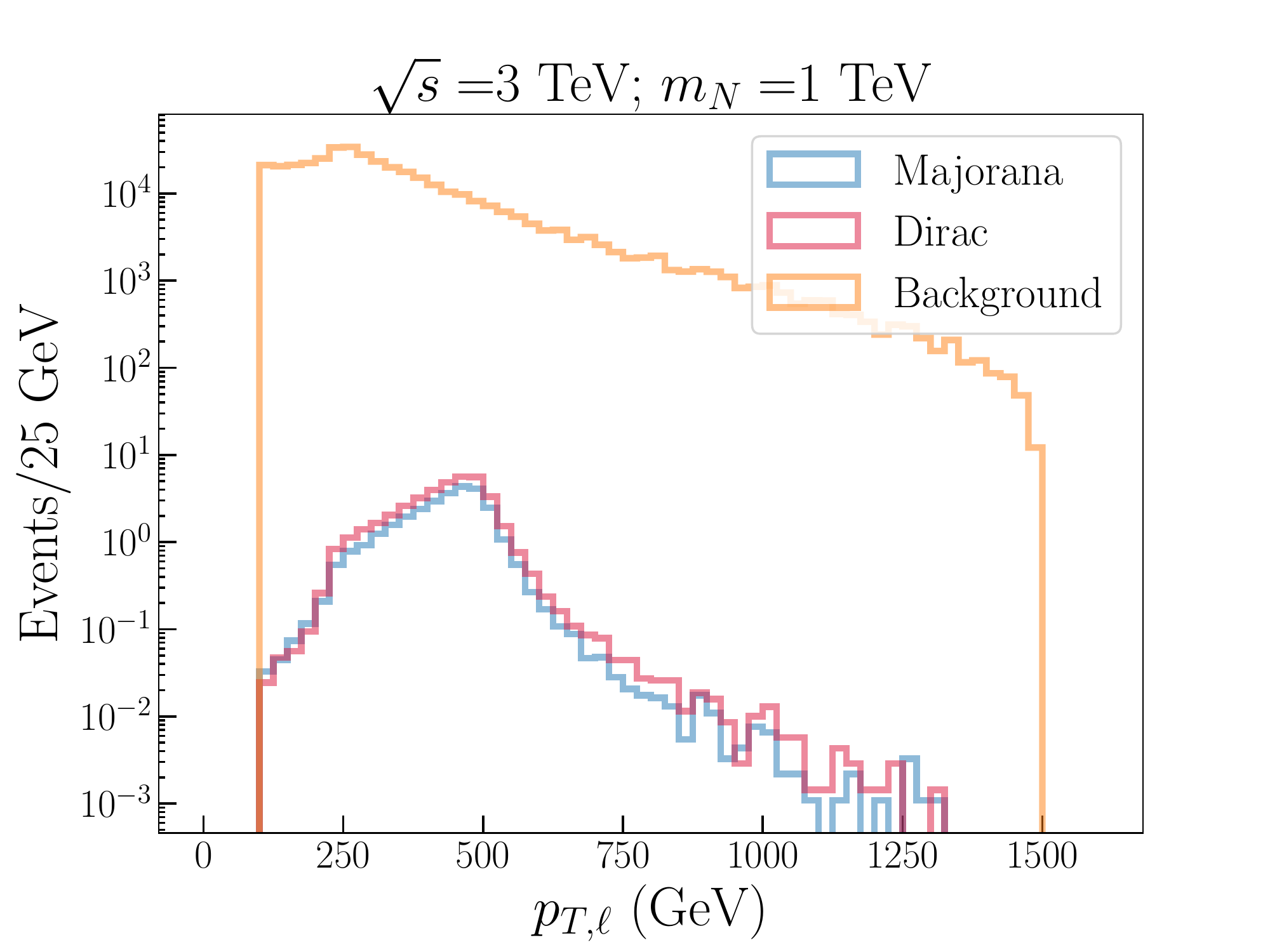}}
    \subfloat[]{\includegraphics[width=0.45\linewidth]{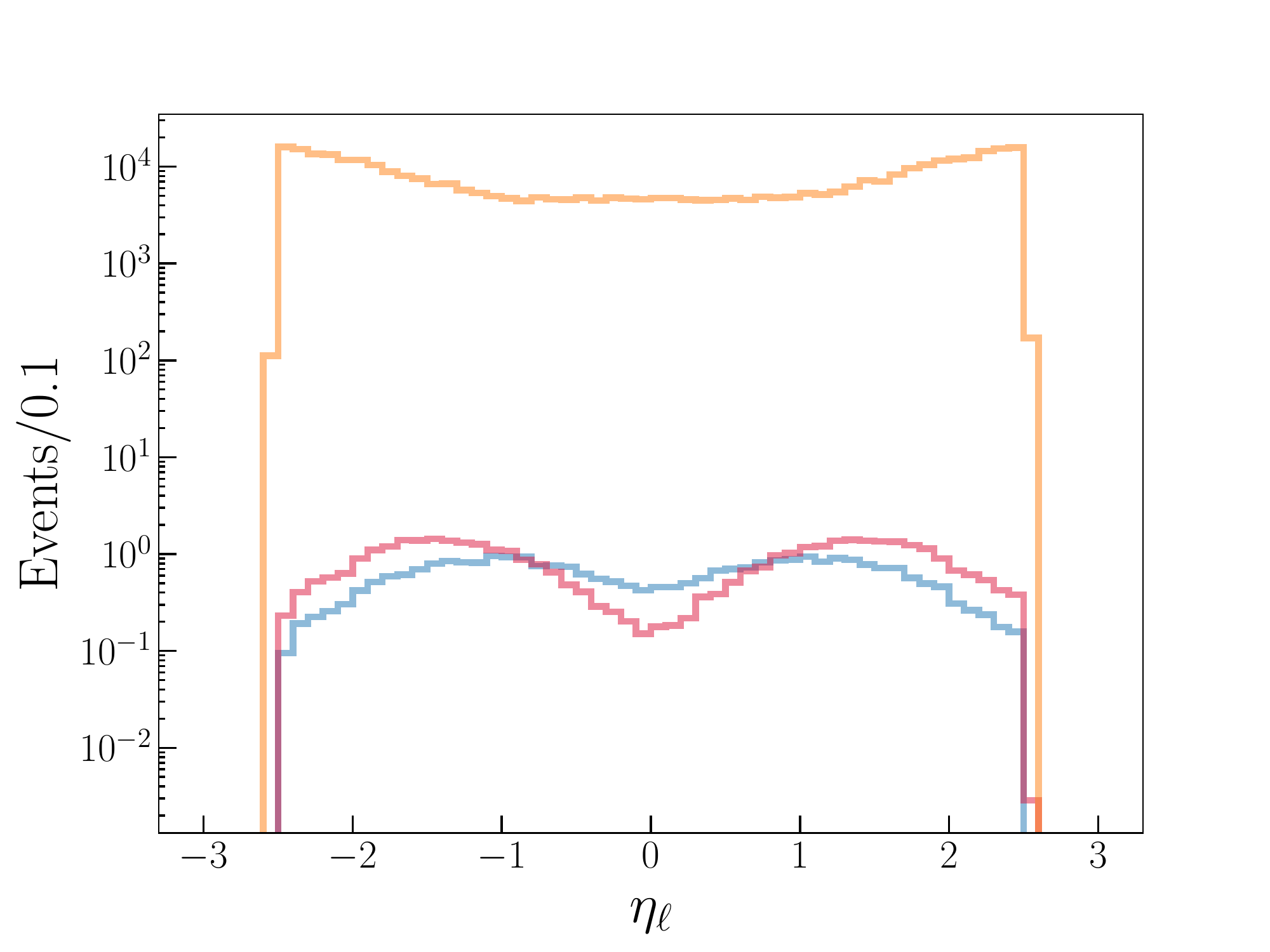}}
    \hfill
    \subfloat[]{\includegraphics[width=0.45\linewidth]{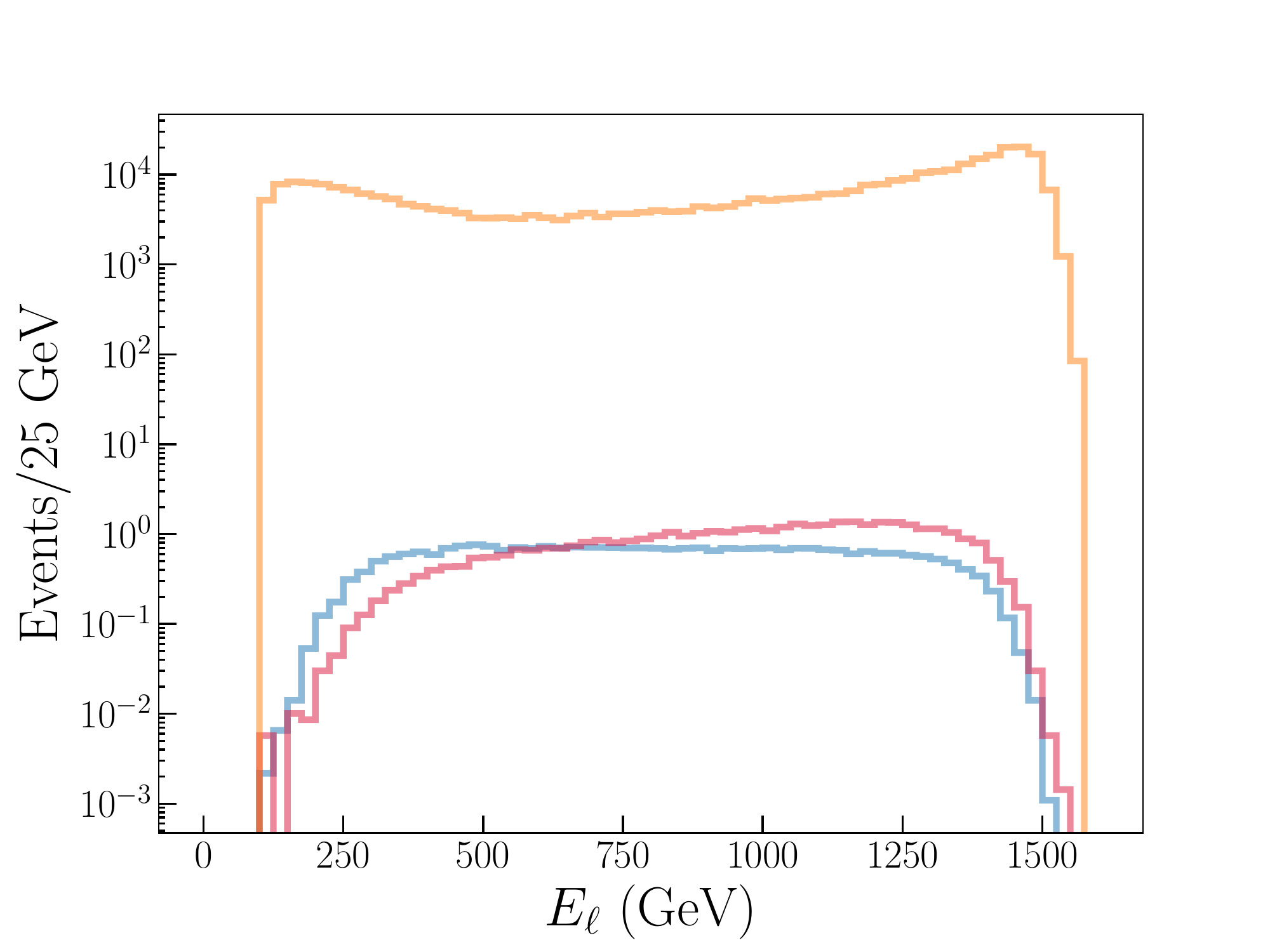}}
    \subfloat[]{\includegraphics[width=0.45\linewidth]{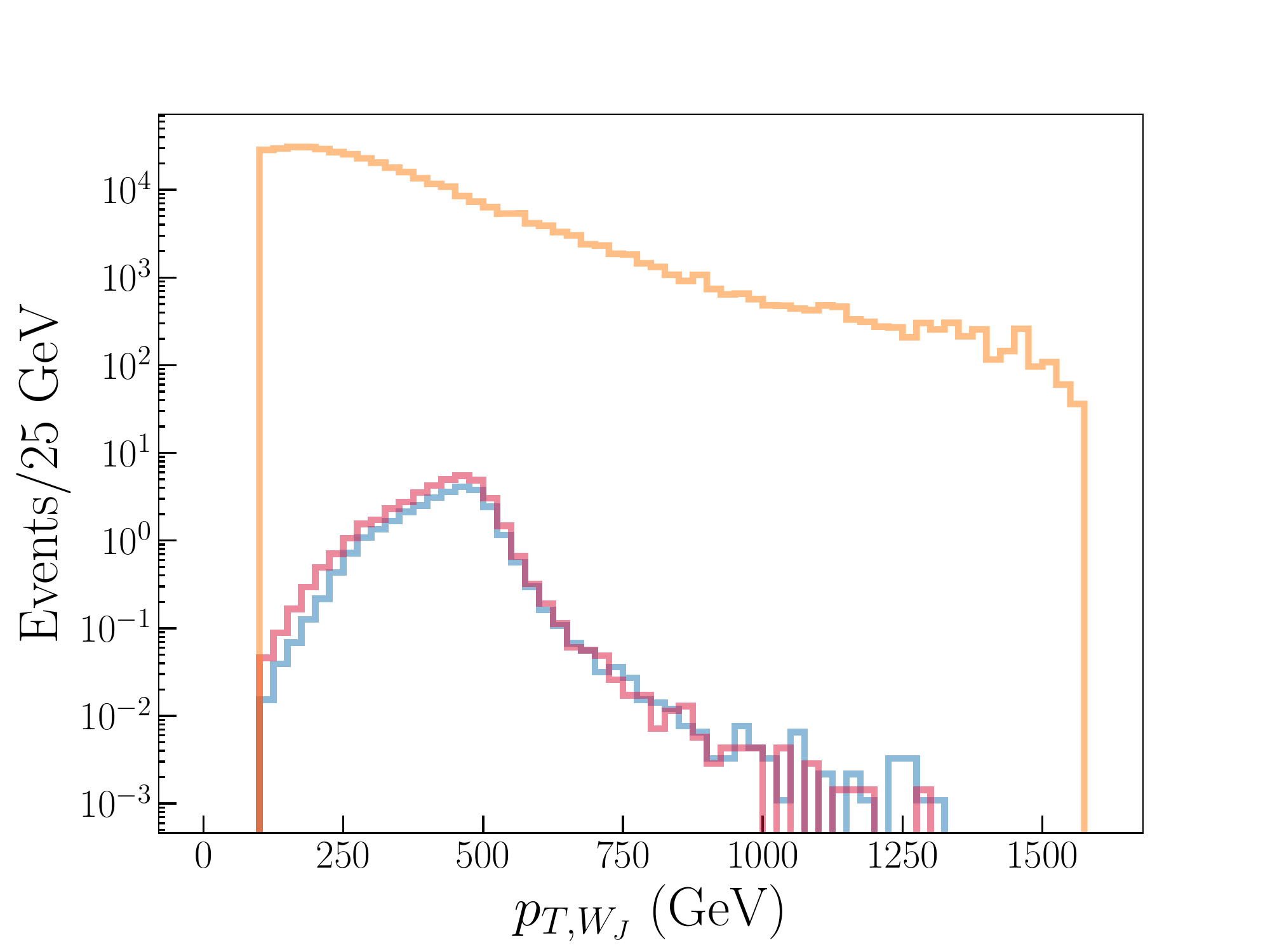}}
    \hfill
    \subfloat[]{\includegraphics[width=0.45\linewidth]{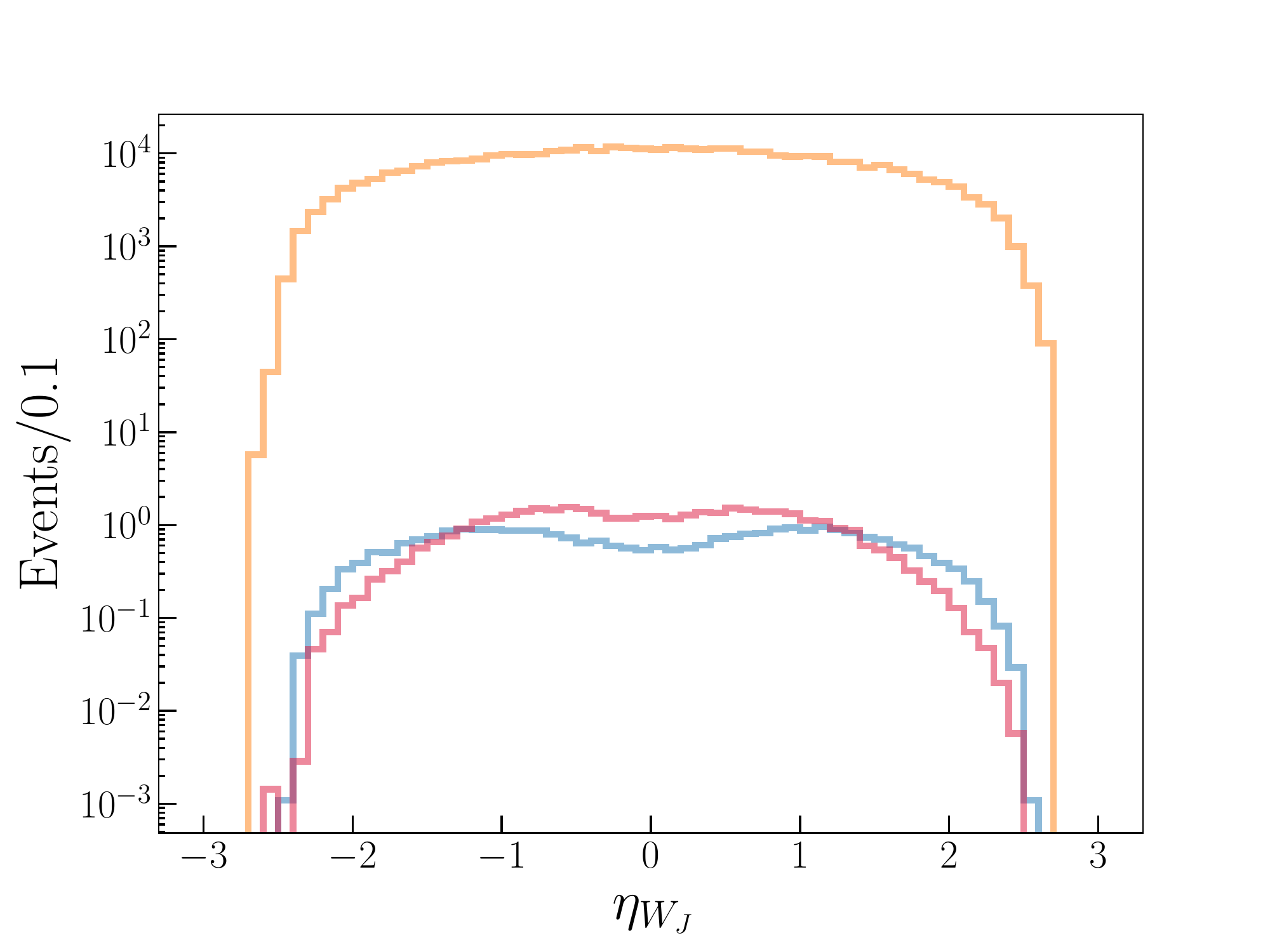}}
    \subfloat[]{\includegraphics[width=0.45\linewidth]{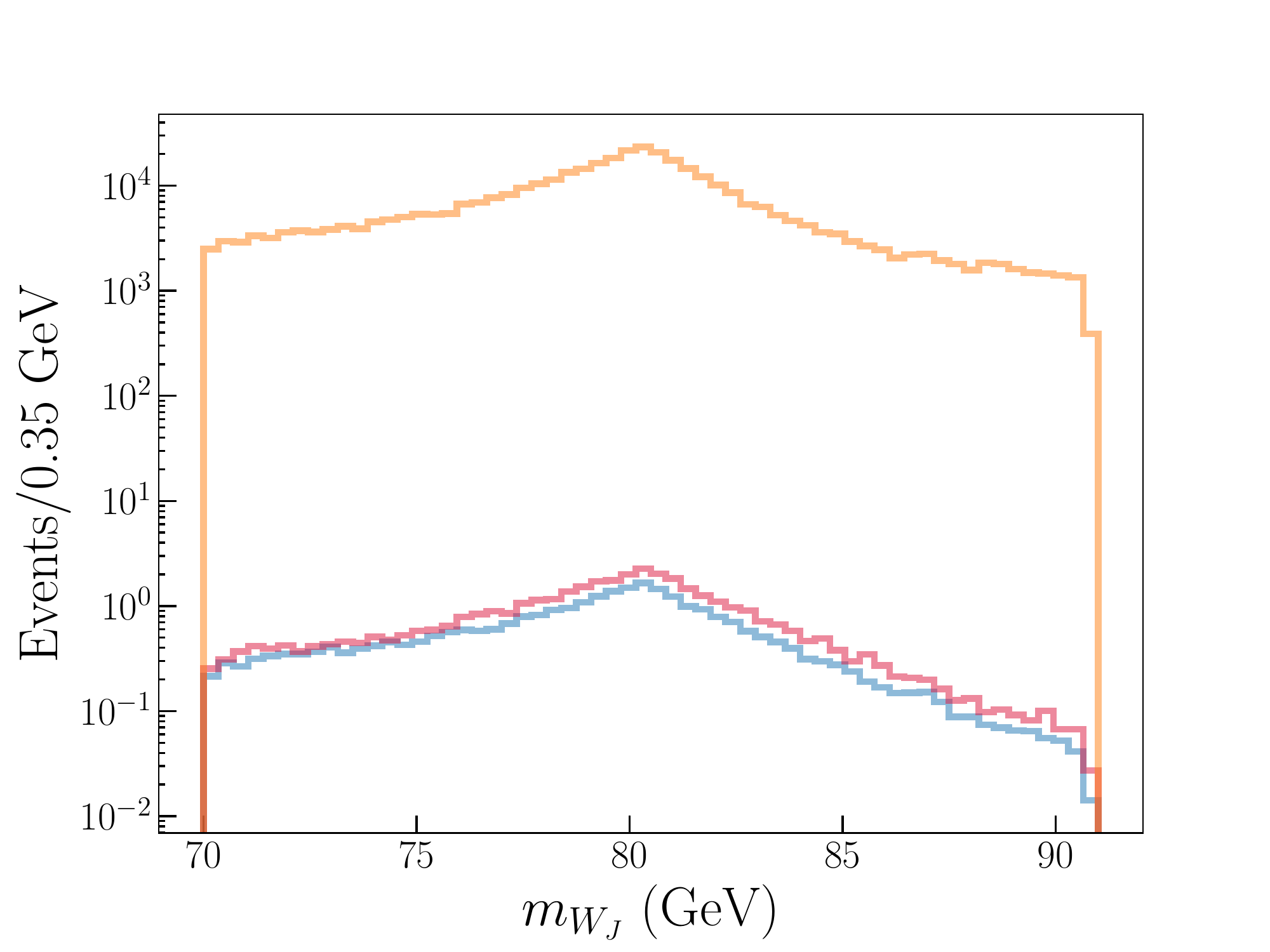}}
    \caption{\label{fig:bdt_features_figs1} Distributions of the features used in the BDT analysis, for a benchmark scenario of $m_N=1$~TeV in a $\sqrt{s}=3$~TeV, $L=1$~ab$^{-1}$ muon  collider. Blue corresponds to Majorana signal events, red for Dirac, and yellow for the combined SM background. For all signal events, cross sections are normalized to their $95\%$ exclusion limits. }
\end{figure}

\begin{figure}[htb!]
    \centering
    \subfloat[]{\includegraphics[width=0.45\linewidth]{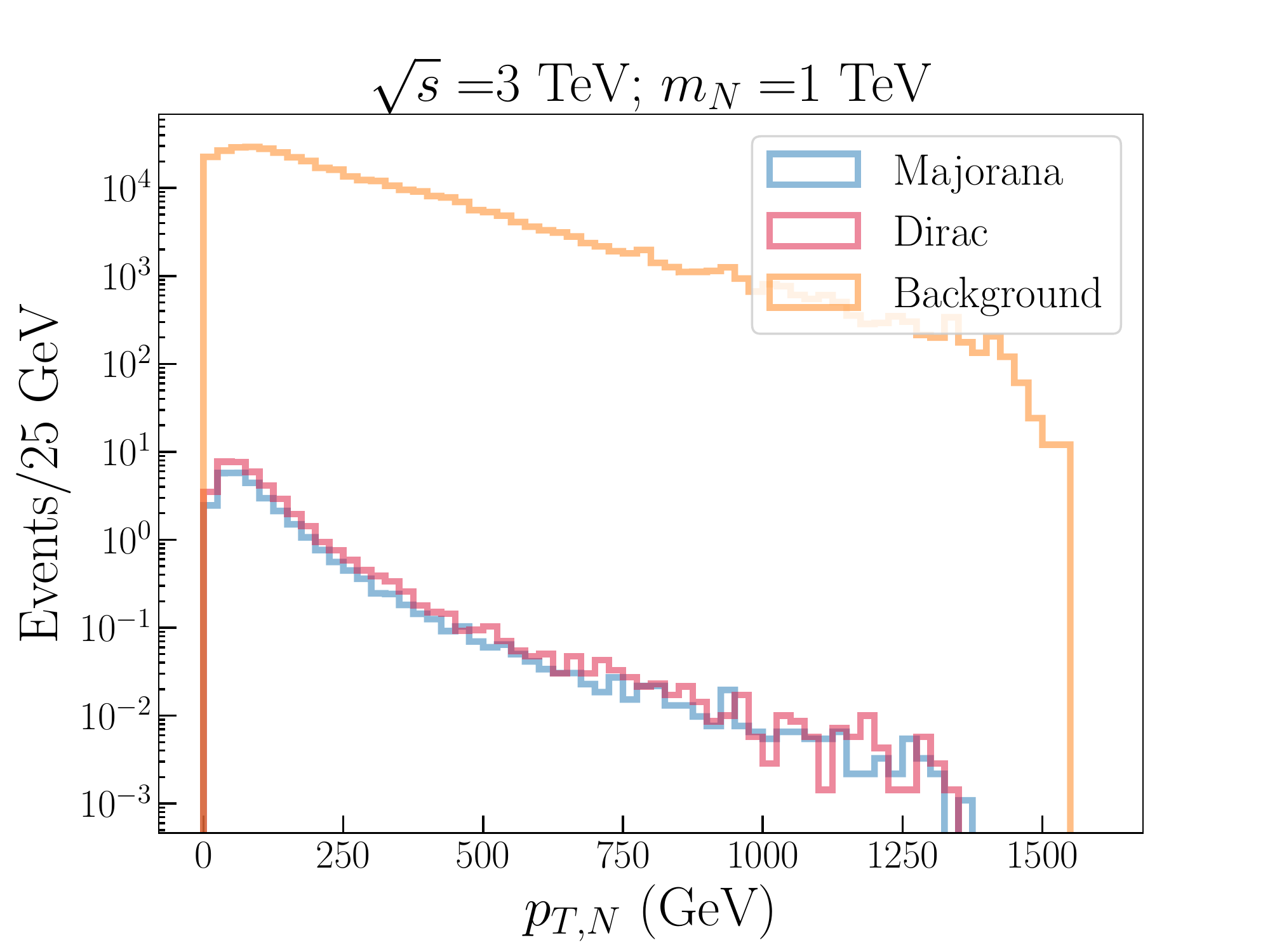}}
    \subfloat[]{\includegraphics[width=0.45\linewidth]{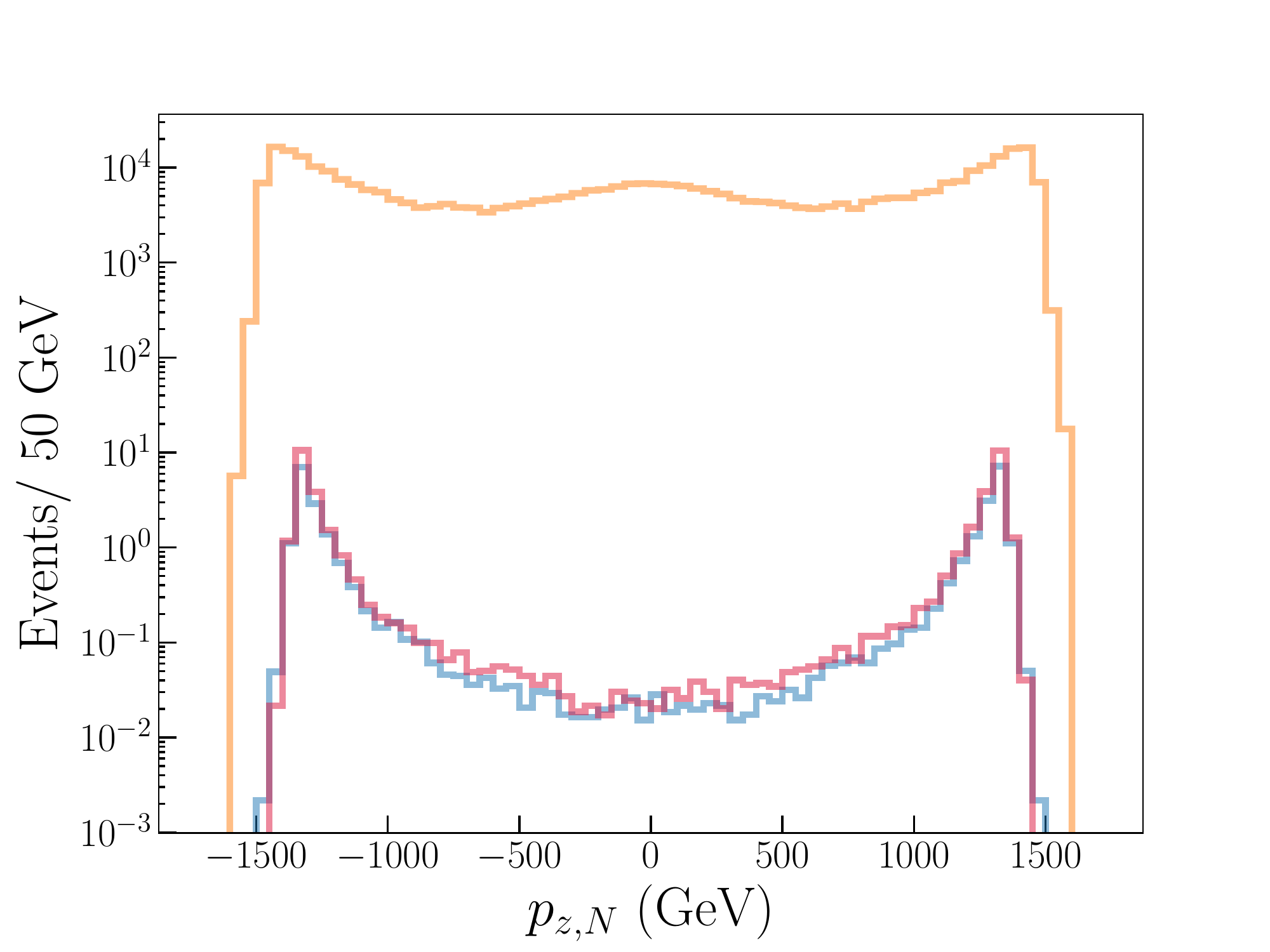}}
    \hfill
    \subfloat[]{\includegraphics[width=0.45\linewidth]{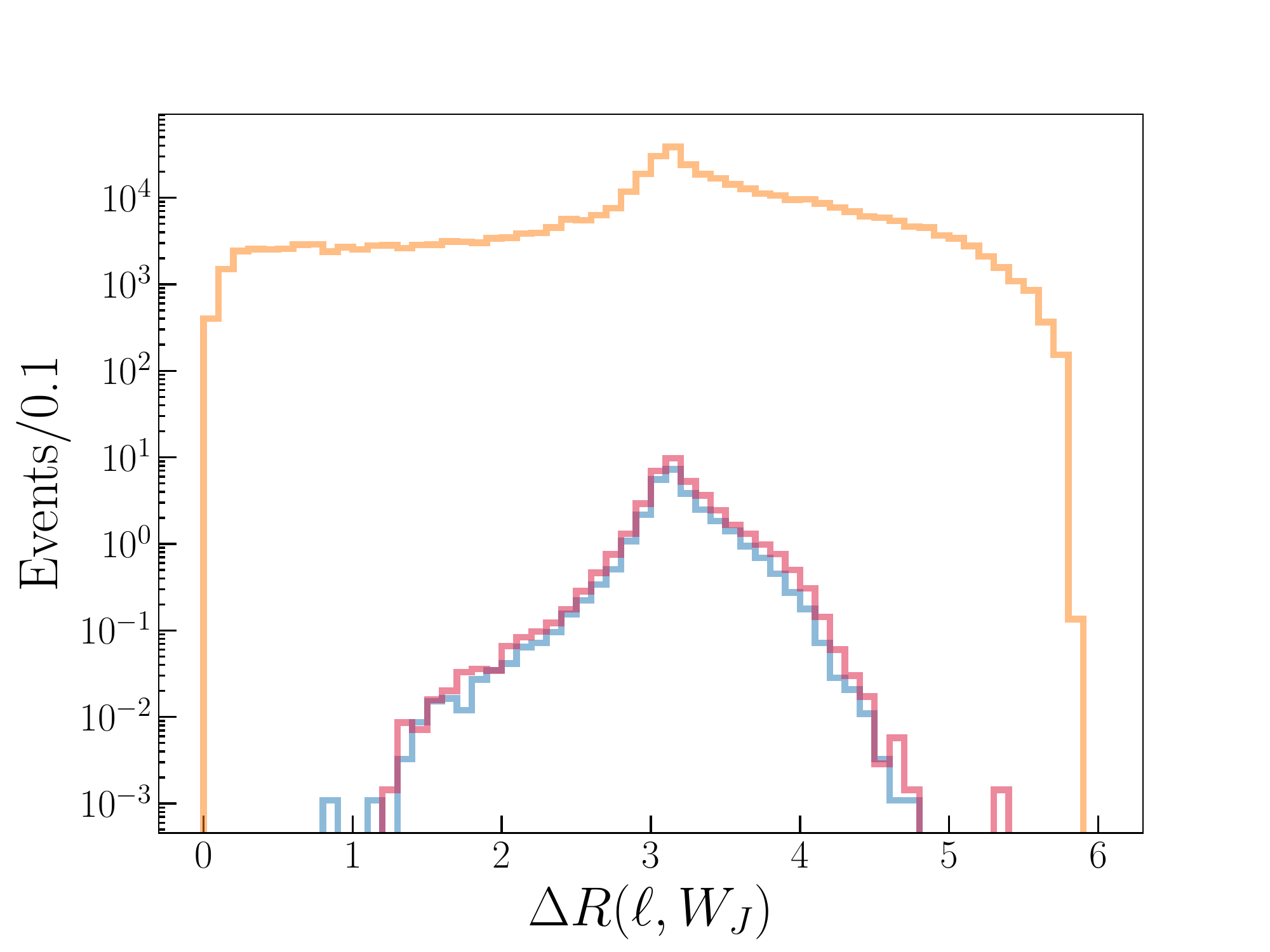}}
    \subfloat[]{\includegraphics[width=0.45\linewidth]{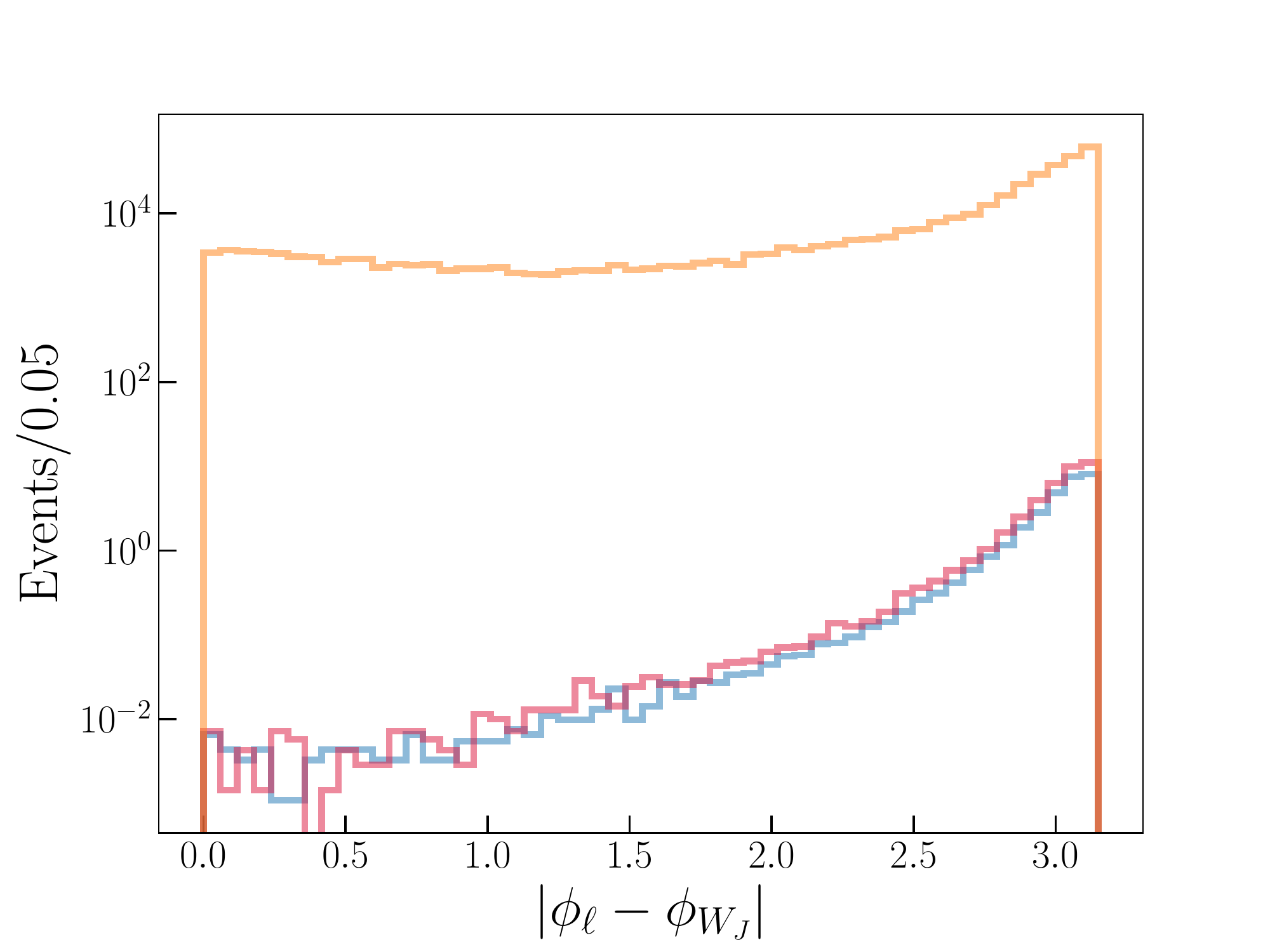}}
    \hfill
    \subfloat[]{\includegraphics[width=0.45\linewidth]{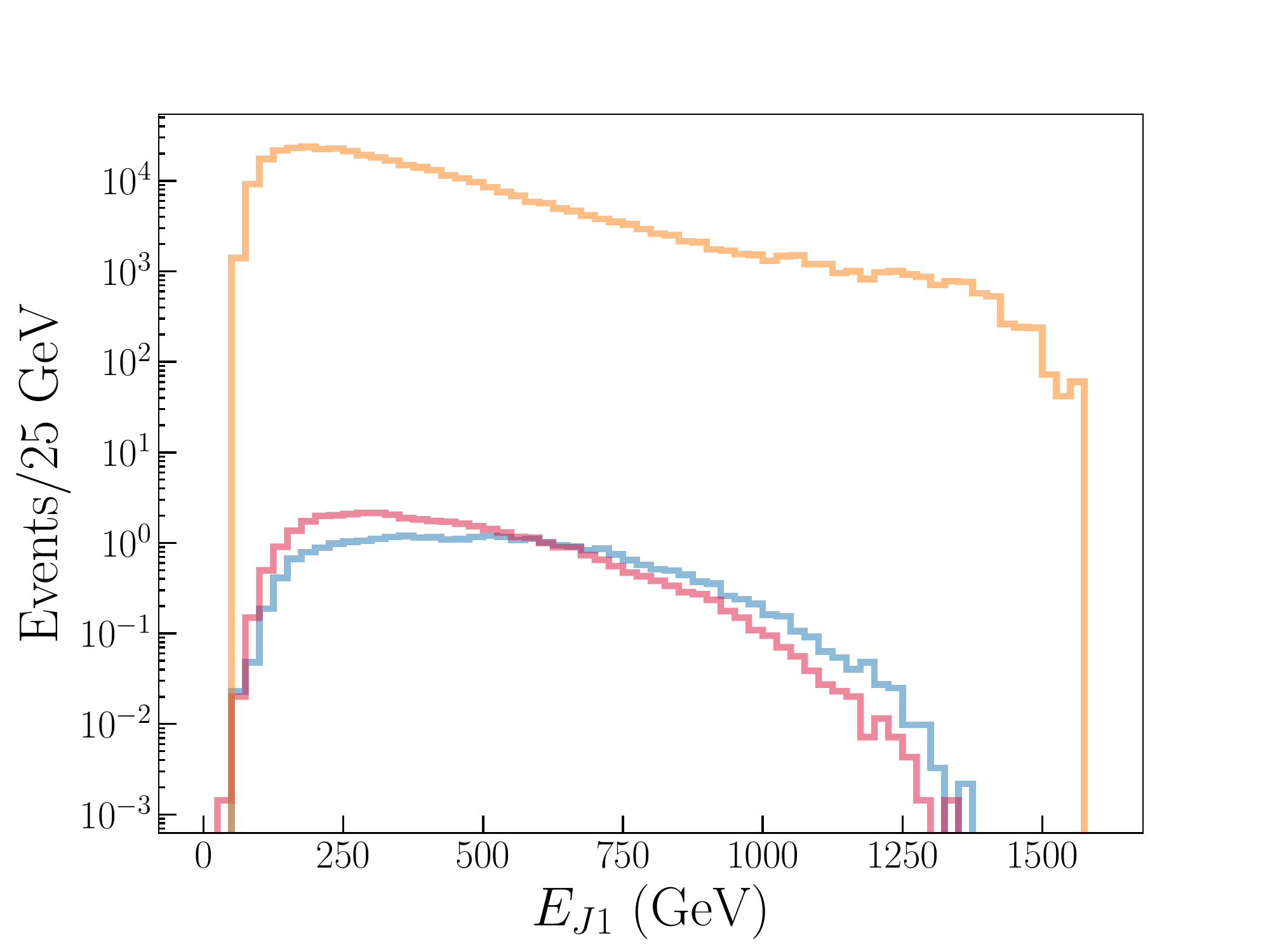}}
    \subfloat[]{\includegraphics[width=0.45\linewidth]{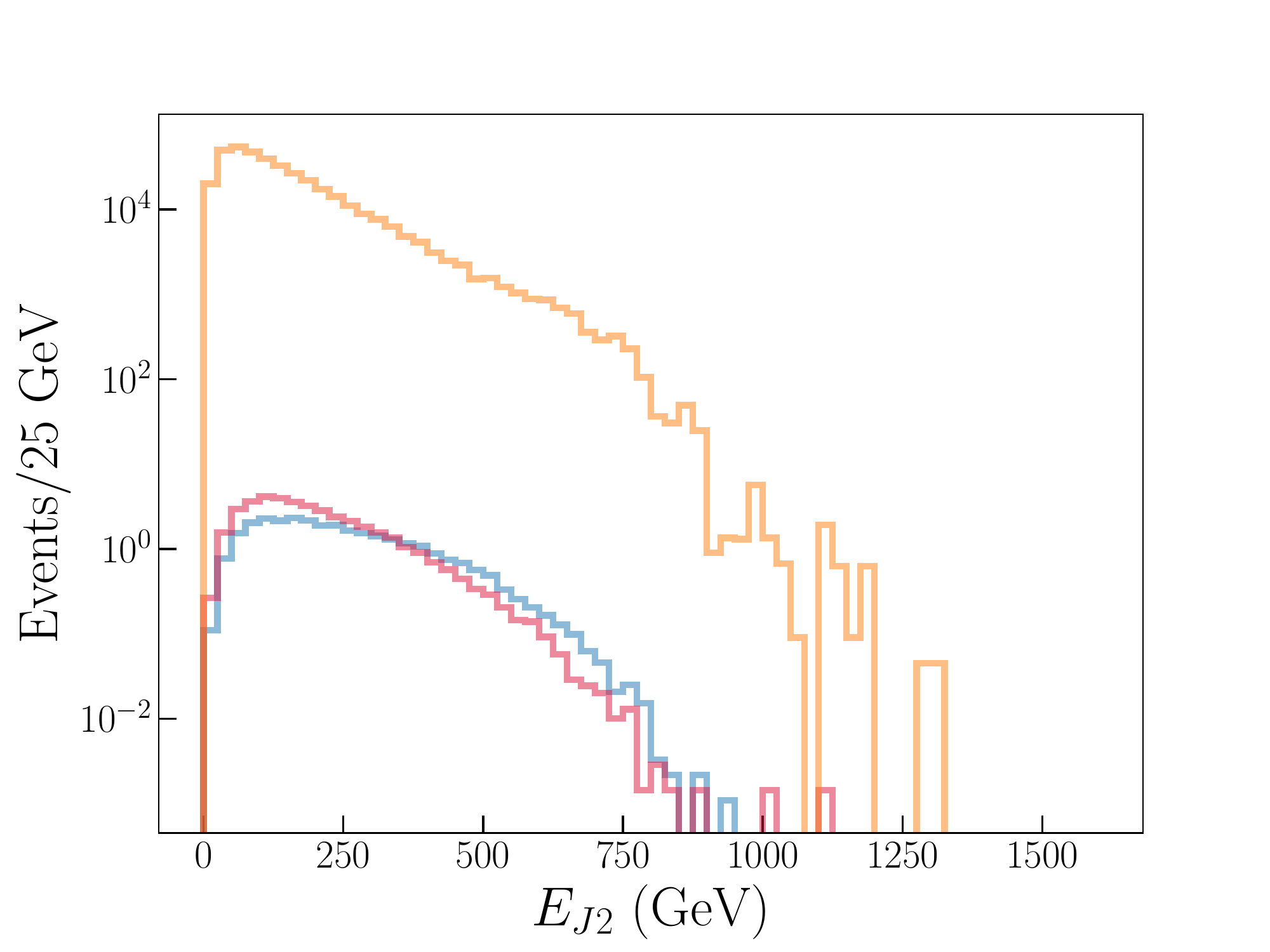}}
    \caption{\label{fig:bdt_features_figs2} Continuation of figure~\ref{fig:bdt_features_figs1}. Distributions of the features used in the BDT analysis, for a benchmark scenario of $m_N=1$~TeV in a $\sqrt{s}=3$~TeV, $L=1$~ab$^{-1}$ muon  collider. Blue corresponds to Majorana signal events, red for Dirac, and yellow for the combined SM background. For all signal events, cross sections are normalized to their $95\%$ exclusion limits.}
\end{figure}

We show the distributions for the BDT features in figures~\ref{fig:bdt_features_figs1} and~\ref{fig:bdt_features_figs2}. Generally, for most variables, the kinematic distributions for Majorana and Dirac HNLs are similar, but the distributions of the energy and pseudorapidity of the lepton $\eta_\ell$ (figure~\ref{fig:bdt_features_figs1}~(b)) and $E_\ell$ (figure~\ref{fig:bdt_features_figs1}~(c)), and of the pseudorapidity of the jet system $\eta_{W_J}$  (figure~\ref{fig:bdt_features_figs1}~(e)) differ for the two types of HNL. The flatness of the distribution in $E_\ell$ for Majorana HNLs is due to the lack of forward-backward asymmetry in their decay. More broadly,  This effect is well known~\cite{Petcov:1984nf,BahaBalantekin:2018ppj,Balantekin:2018ukw,deGouvea:2021ual,deGouvea:2021rpa} and has been noted in the context of hadron~\cite{Arbelaez:2017zqq} and lepton~\cite{Hernandez:2018cgc,Mekala:2022cmm} colliders. Additionally, since the $p_T$ distributions for Majorana and Dirac HNLs are functionally identical, it then follows that pseudorapidity distributions vary between the two types of HNLs in an analogous fashion to the $E_\ell$ distributions.


\begin{figure}[tbp]
\centering
\includegraphics[width=0.65\linewidth]{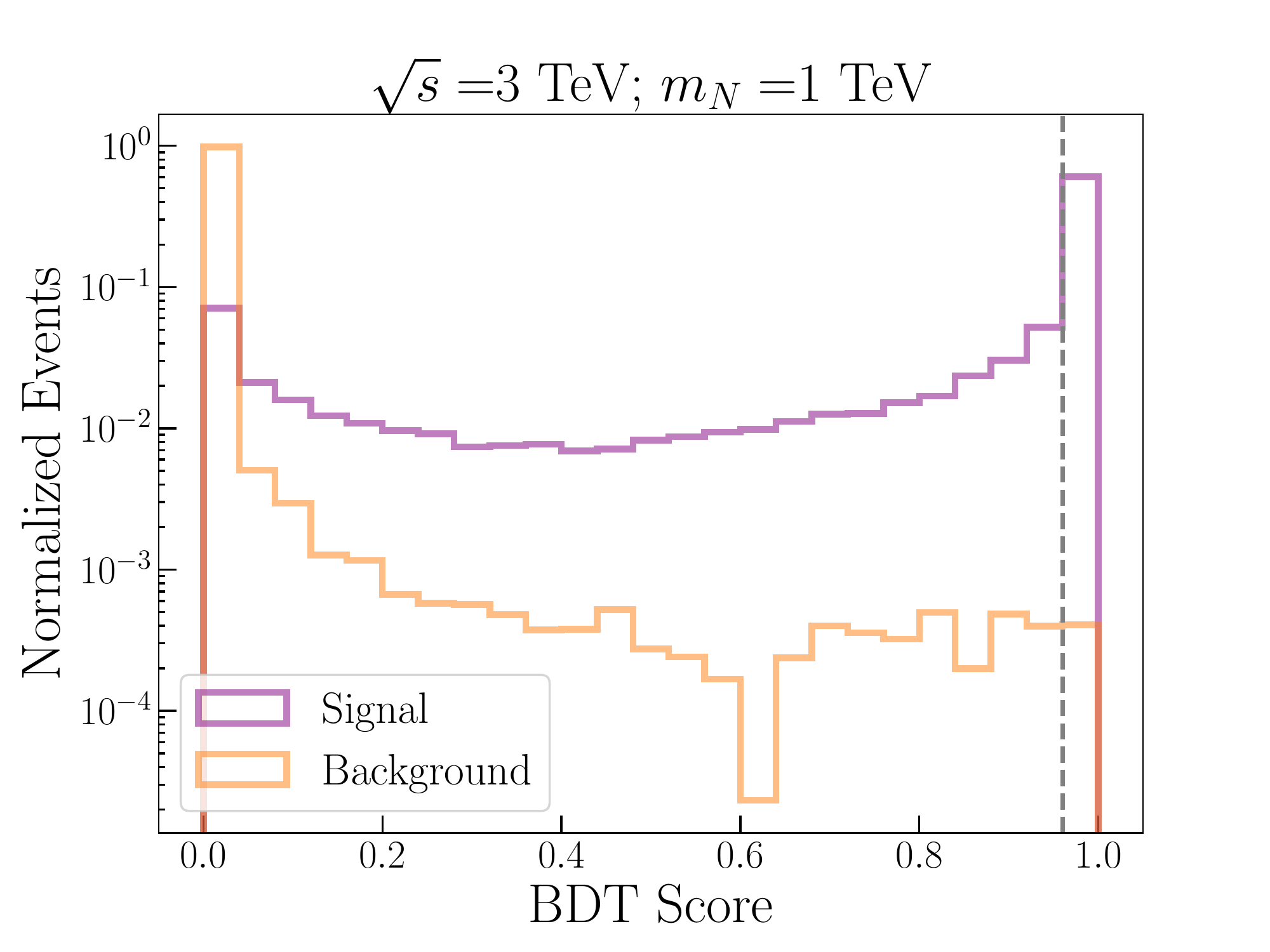}
\caption{The normalized BDT response, $1-P_B$, for a benchmark scenario of $m_N=1$~TeV in a $\sqrt{s}=3$~TeV, $L=1$~ab$^{-1}$ muon collider. The events in yellow correspond to the combined SM background, and the purple events correspond to the combined the Majorana and Dirac signals. The dashed vertical line is the cut on the BDT score that we use to distinguish signal from background. The optimization of the BDT cut here is operated on a grid four times finer than the binning shown in the plot.}
\label{fig:benchmarkTchanBDTscore}
\end{figure}

As we run the BDT analysis as a three-class categorization task, we use the quantity $1-P_B$ as the BDT response score, where $P_B$ is the probability the algorithm classifies an event as background. The BDT response for a benchmark scenario of $m_N=1$~TeV in a $\sqrt{s}=3$~TeV collider is shown in figure~\ref{fig:benchmarkTchanBDTscore}. The dashed vertical line in figure~\ref{fig:benchmarkTchanBDTscore} is the cutoff on the BDT score that we use to separate signal from background. In the analysis of section~\ref{subsec:v2_exc}, only the $1-P_B$ cut is applied. For use in the analysis of section \ref{subsec:majvsdir}, on events that pass the background cut we also apply a cutoff on the quantity $P_M-P_D$ to separate Majorana-like and Dirac-like signal events, where $P_M$ ($P_D$) is the probability the BDT algorithm classifies the signal as Majorana (Dirac). We optimize both the $1-P_B$ and $P_M-P_D$ cuts simultaneously.

\begin{figure}[tbp]
\centering
\includegraphics[width=0.65\linewidth]{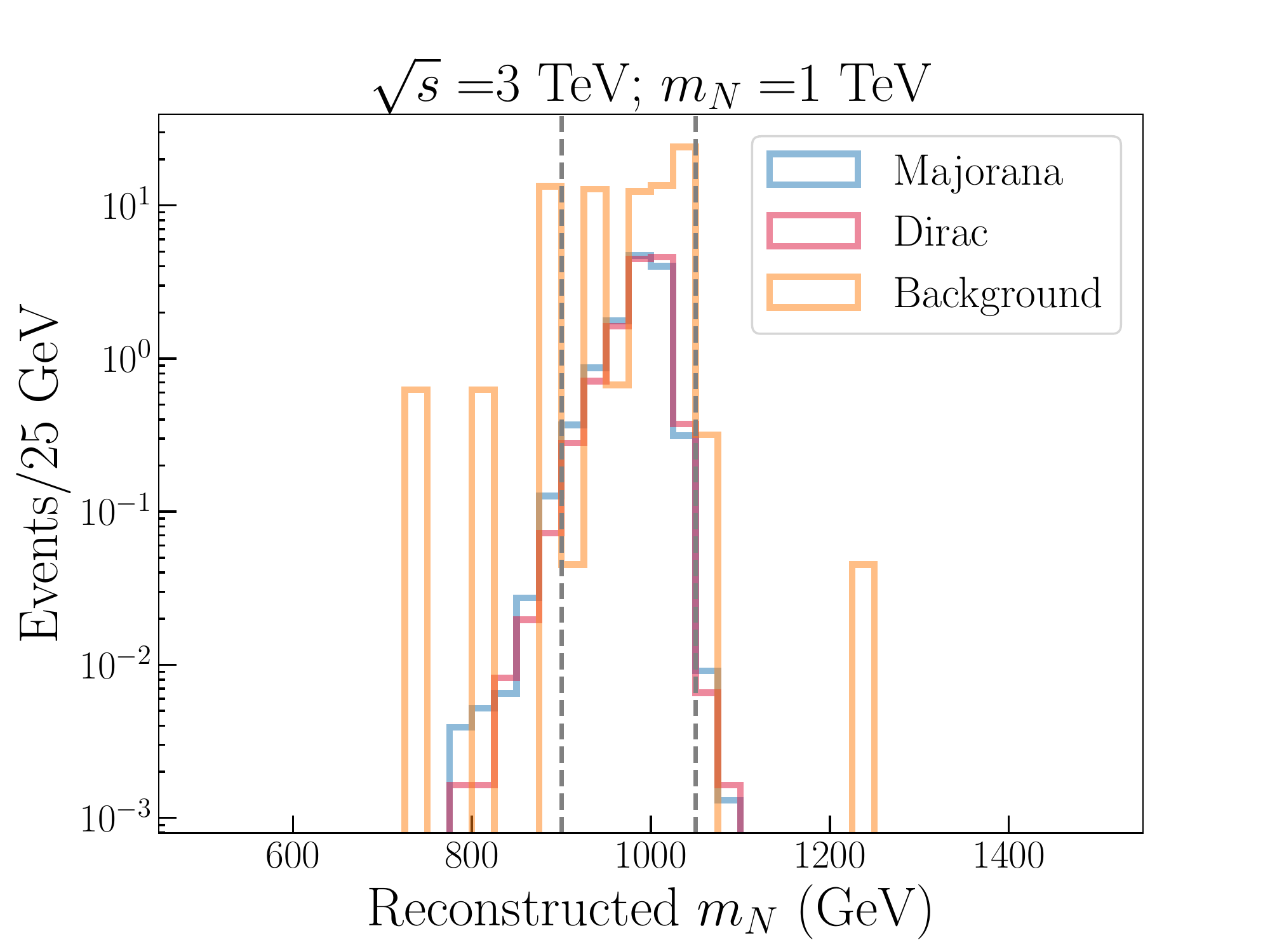}
\caption{A histogram of reconstructed $m_N$ after applying BDT, for a benchmark scenario of $m_N=1$~TeV in a $\sqrt{s}=3$~TeV, $L=1$~ab$^{-1}$ muon collider. The dashed vertical lines are the cuts on the $m_N$ mass used to further separate signal from background. Blue corresponds to Majorana signal events, red for Dirac, and yellow for the combined SM background. For all signal events, cross sections are normalized to their $95\%$ exclusion limits.}
\label{fig:benchmarkTchanMassHist}
\end{figure}

In addition to cutting on the BDT score, we further apply a cut on the mass of the reconstructed HNL, namely within $[0.9\times m_N, 1.05\times m_N]$. This window is asymmetric, as the reconstructed distribution has a low mass tail.  Figure~\ref{fig:benchmarkTchanMassHist} shows the invariant mass distribution $m_N$ of the HNL signals and SM background after the application of the BDT analysis.

\section{\label{sec:Results}Results}
\subsection{\label{subsec:v2_exc}Exclusion Limits on $|V_\ell|^2$}

We now present the results of the BDT analysis in order to construct $95\%$ exclusion limits on the HNL mixing parameter $|V_\ell|^2$. 
At this level of analysis, we assume no systematic uncertainty in the number of recorded events. Our goal is to determine exclusion limits at $95\%$ confidence, and as such we require $Z\geq 1.96$. Therefore, for each benchmark simulation we run the BDT analysis and extract the lower limit on $|V_\ell|^2$ such that the signal significance is above threshold. This analysis was done using a three-class BDT. We applied a two-class analysis to cross check and found exclusion curves with comparable sensitivies.

\begin{figure}[tb]
\centering
\includegraphics[width=\linewidth]{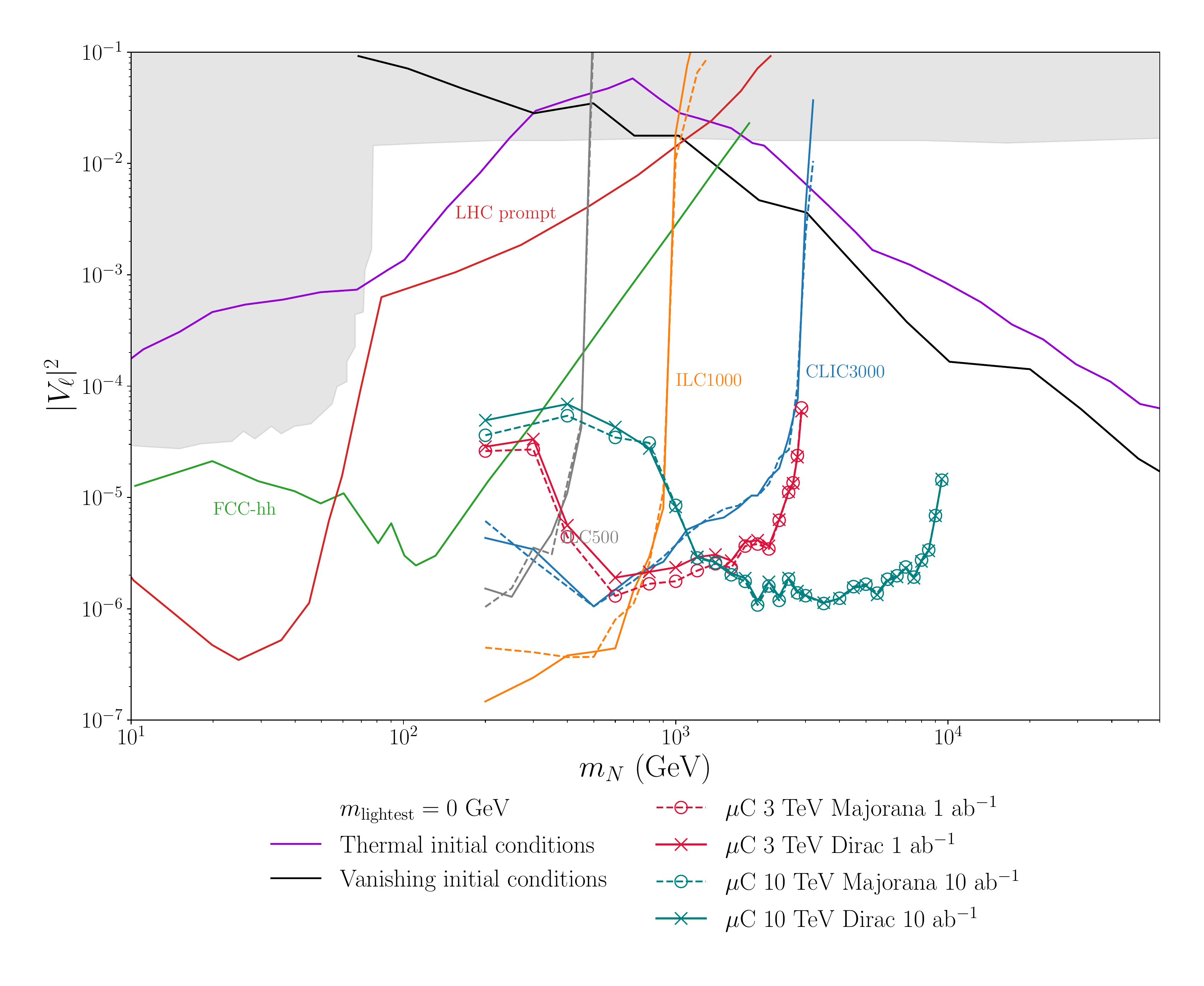}
\caption{\label{fig:results} $95\%$ exclusion limits for $|V_\ell|^2=|V_e|^2=|V_\mu|^2$ as a function of the HNL mass $m_N$. The red dashed (solid) line corresponds to a Majorana (Dirac) HNL at a muon collider with $\sqrt{s}=3$~TeV, $L=1$~ab$^{-1}$. The teal dashed (solid) line corresponds to a Majorana (Dirac) HNL at a muon collider with $\sqrt{s}=10$~TeV, $L=10$~ab$^{-1}$. The solid black and purple lines correspond to limits from considerations of viable leptogenesis scenarios~\cite{Drewes:2021nqr}. The grey area is the region excluded by a global scan~\cite{Chrzaszcz:2019inj}. The red line shows the limits from prompt trilepton searches at the LHC ~\cite{Izaguirre:2015pga}. The green line shows the limits from a future FCC-$hh$~\cite{Antusch:2016ejd}. Lastly, the grey, orange and blue lines are exclusion limits in future $e^+e^-$ linear colliders~\cite{Mekala:2022cmm}.}
\end{figure}

The results of this process are shown in figure~\ref{fig:results}, as well as comparisons to similar searches in other (proposed) experiments. For masses above $m_N>1$~TeV, the exclusion limits are stronger than any other limits shown so far, with reach down to $|V_\ell|^2\sim \mathcal{O}(10^{-6})$ or even lower. Note that in this mass region, it is often the case that the sensitivities for the Majorana signals are slightly better than those of the Dirac signals. The reason is evident from figure~\ref{fig:bdt_features_figs1} (c, e), and figure~\ref{fig:bdt_features_figs2} (e, f), which show that the corresponding distributions of a Majorana HNL differ from the SM background distributions more significantly than those of a Dirac HNL.

In the region of lighter HNL masses, namely $m_N<1$~TeV, our limits are generally less strict than those for other experiments, due to the poor reconstruction efficiency for low mass HNLs. Nevertheless, the uncertainty in the signal yield for these benchmarks is high, and as such our limits can be considered conservative in this region. 

Furthermore, in this low $m_N$ region, the HNL signal would be dominated by the VBF channel, wherein two vector bosons from the muon beams give rise to $N+\ell$ or $N+\nu$ final states. The reconstruction efficiency, in this case, could be significantly larger than that of the $t$-channel ones, as the $p_z$ of the HNL produced would be moderate, leaving more decay products in the central region. In addition, in the scenario where $|V_\mu|\ll |V_e|$ or $|V_\tau|$, the $t$-channel signal rate would be greatly suppressed, whereas the VBF channel signal rate could still be relevant. Nevertheless, the biggest problem for the VBF channel is that its production rate is at least two orders of magnitude smaller than that of the $t$-channel process, since the process of higher order. Consequently, for this work's benchmark of $|V_\ell|=|V_e|=|V_\mu|$, the estimated exclusion limit of $|V_\ell|$ is at most comparable to that of the $t$-channel when $m_N\simeq 200$~GeV, and drops much faster as $m_N$ increases. We will leave a through study of VBF channels to future work.
\subsection{\label{subsec:majvsdir}Distinguishing Majorana versus Dirac Heavy Neutral Leptons}
\begin{figure}[tb]
    \centering
    \includegraphics[width=\linewidth]{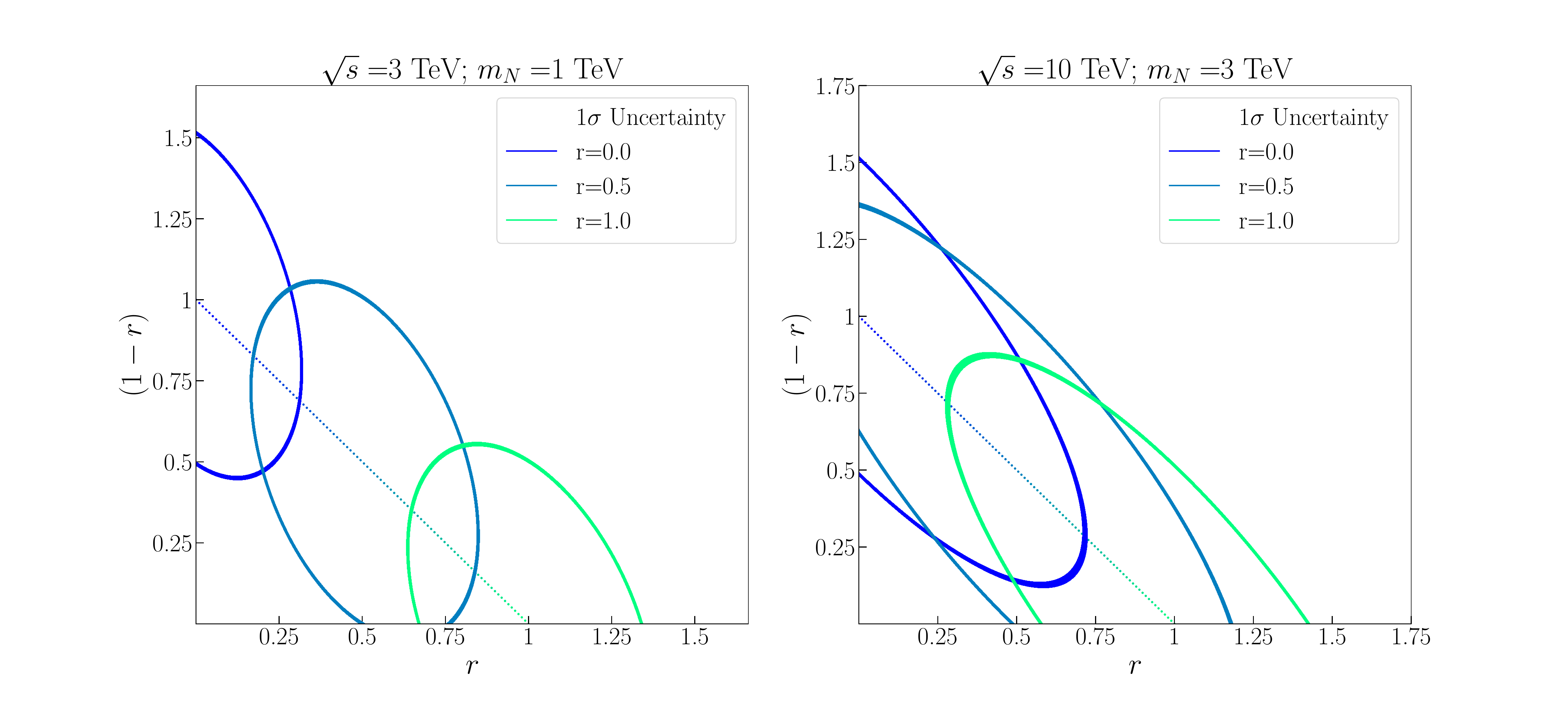}
    \caption{The two-dimensional likelihood in the plane of Majorana and Dirac HNL signal yields in a $\sqrt{s}=3$~TeV (left) and a $\sqrt{s}=10$~TeV collider (right). The 3 ellipses on each panel represent the $1\sigma$ contours centered at $r=0,\,0.5$ and $1.0$.  Note that the ellipses are tilted, and showing a negative correlation between the measured Majorana and Dirac signal yields. All contours are evaluated at their $95\%$ exclusion limits where $|V_\ell|^2 \sim\mathcal{O}(10^{-6})$, respectively.}
    \label{fig:combined_uncertainty}
\end{figure}

As noted in section~\ref{sec:Analysis} and figure~\ref{fig:bdt_features_figs1}, there are signatures in the final state kinematic distributions that differ between Majorana and Dirac type HNLs. It is therefore an interesting question to ask what would be the discrimination potential to distinguish Majorana versus Dirac HNLs in a muon collider. We consider the quantity $r=n_M/(n_M+n_D)$, where $n_M$ ($n_D$) is the number of detected Majorana (Dirac) HNLs. Note that this definition differs from the usual $R_{ll}$ seen in the literature, which is based on the ratio of lepton number violating (LNV) decays to lepton number conserving (LNC) decays (see, e.g.~\cite{Drewes:2022rsk}).\footnote{The two definitions agree for $r=R_{ll}=0,1$, but our $r=1/2$ corresponds to $R_{ll}=1/3$. Alternatively, $r$ could also be defined as $\frac{2n_{\rm LNV}}{n_{\rm{LNC}} +n_{\rm{LNV}} }$, where $n_{\rm LNC}$ and $n_{\rm LNV}$ are numbers of detected LNC and LNV decays, respectively.} In order to quantify our analysis, we consider the two-dimensional likelihood in the plane of Majorana and Dirac HNL signal yields in a $\sqrt{s}=3$~TeV and a $\sqrt{s}=10$~TeV collider, shown in figure~\ref{fig:combined_uncertainty} on the left and right-hand sides respectively. Both the horizontal and vertical axes are normalized by the same quantity, namely the $95\%$ exclusion limits $|V_\ell|^2$ as given in figure~\ref{fig:results} for either Majorana or Dirac type --- whichever value is more conservative. Each point on the the dotted line represents a possible value of $r$, and the $x$ and $y$ intercepts of that line correspond to the fully Majorana ($r=1$) and Dirac ($r=0$) cases respectively. For a given $r$ value, we can construct the corresponding $1\sigma$ likelihood contour based on the Majorana and Dirac regions defined in section~\ref{sec:Analysis}. In this analysis, a signal event is classified as Majorana or Dirac if it appropriately satisfies the $P_M-P_D$ cut correctly. If not, it is counted as additional background. 

We plot three such contours at $r=0,\,0.5$ and $1.0$. The centers of these ellipses lie on the diagonal dotted line, and the boundaries of each correspond to fluctuations in signal yields with significance $1\sigma$. Note that the ellipses are tilted, and showing a negative correlation between Majorana and Dirac signal yields. This is due to the fact that the classification is imperfect, as some Majorana or Dirac decays could be misclassified as each other while keeping total HNL signal rate the same. The contours for the $\sqrt{s}=10$~TeV benchmark are tilted more strongly (i.e more highly correlated). This indicates that a downwards fluctuation of Majorana HNL decay is more likely to be due to misclassification as a Dirac HNL than misclassification as background, as compared to the case in the $\sqrt{s}=3$~TeV benchmark. 

\begin{figure}[tb]
    \centering
    \includegraphics[width=\linewidth]{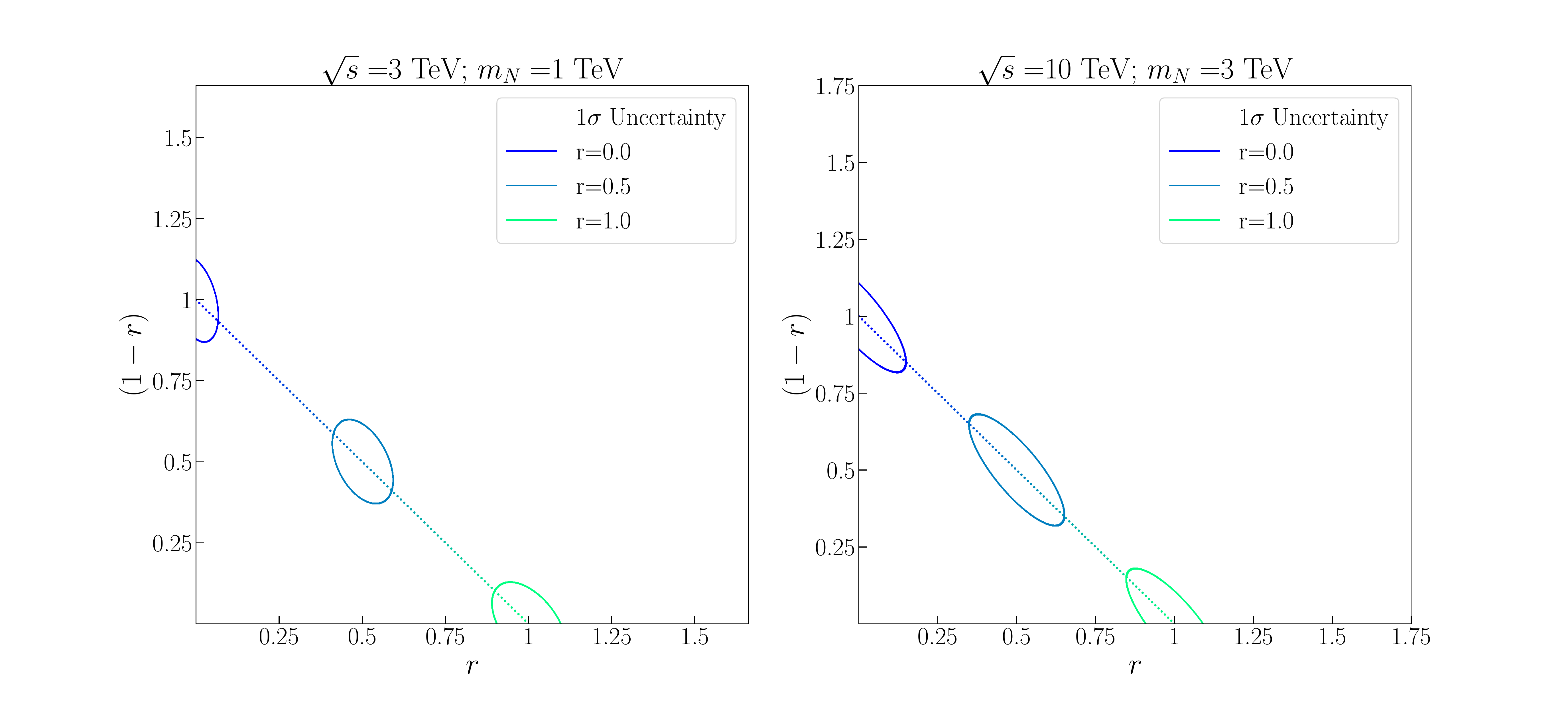}
    \caption{The two-dimensional likelihood in the plane of Majorana and Dirac HNL signal yields in a $\sqrt{s}=3$~TeV (left) and a $\sqrt{s}=10$~TeV collider (right). The 3 ellipses on each panel represent the $1\sigma$ contours centered at $r=0,\,0.5$ and $1.0$.  Note that the ellipses are tilted, and showing a negative correlation between Majorana and Dirac signal yields. The $|V_l|^2$ used here are five times larger than their counterparts in figure~\ref{fig:combined_uncertainty}, leading to much sharper contours.}
    \label{fig:combined_uncertainty_large}
\end{figure}

Lastly, the uncertainties shown in figure~\ref{fig:combined_uncertainty} are marginal, as we are considering discrimination potential at the $95\%$ exclusion limits. We have found that repeating this analysis for larger $|V_\ell|^2$ results in much tighter contours since the larger statistics improve the precision of both signal rates. For example, in figure~\ref{fig:combined_uncertainty_large} we repeat the above analysis with $|V_\ell|^2=5\times |V_{\ell}|^2_{95\%}$. Note that the values for here $|V_\ell|^2$ are still small --- of order $\mathcal{O}(10^{-5}).$  We see that we therefore have a nontrivial discrimination potential for the HNL's type as long as its signal yield is above the exclusion limit. 

\section{\label{sec:Conclusion}Conclusion and Future Directions}
In this paper we investigate the potential for searching for HNLs in a future high-energy muon collider. We consider an effective theory benchmark that couples a single HNL to the first two generations of active neutrinos with equal mixing. We consider HNLs of both Majorana and Dirac types, with masses ranging from $200$~GeV to $9.5$~TeV and determine the reach in the mixing parameter $|V_{\ell}|^2$ for each mass, in $\sqrt{s}=3$~TeV, $L=1$~ab$^{-1}$ and $\sqrt{s}=10$~TeV, $L=10$~ab$^{-1}$ muon colliders. We find that for HNL masses greater than 1~TeV the limits in $|V_\ell|^2$ are the strictest collider limits yet. We also show that a future muon collider has strong discrimination potential to distinguish between Majorana and Dirac type HNLs. 

In the future, there are a few aspects of this work that could be broadened. Firstly, we focus our analysis on an effective theory in which the HNL signal is dominated by single production. The purpose of this work was to be conservative and model-independent, but different UV completions could have distinct signatures accessible at a muon collider. Secondly, a more realistic analysis would include the production and decay of multiple HNLs. Similarly, this work assumes a benchmark point of enhanced symmetry for the mixing of the HNLs, namely $|V_{1e}|=|V_{1\mu}|\neq0$ and $|V_{1\tau}|=0.$ Consequently, we do not consider the decay of a HNL to a tau in this work. For other benchmarks, the dominant processes in the production of HNLs might differ. Lastly, while we include a fast detector simulation in this work, we do not incorporate the effects of detector-based systematic uncertainties. Likewise, we do not include systemic uncertainties due the beams nor do we incorporate beam spectra. As a spectrum profile for muon beams is not yet available in \textsc{Whizard}, we also leave such refinements for a future work.

\noindent \textbf{Note Added}: At the same time as this work was shared on the arXiv, another preprint on the same topic was also uploaded \cite{Mekala:2023diu}. While the two works' aims are similar, they differ in their methods and analyses. We are also aware of a separate work on a similar topic using a cut-based analysis \cite{Li:2023}, posted shortly after this paper. We acknowledge the authors of that work for their coordination.

\acknowledgments
LL would like to thank Samuel Homiller, Zhen Liu, and Kun-Feng Lyu for valuable discussions. AR would like to thank Adam Lister and Harry Hausner for useful comments. THK, TL and AR are supported partly by the Area of Excellence (AoE) under the Grant No.~AoE/P-404/18-3 and supported partly by the General Research Fund (GRF) under Grant No.~16304321. Both of the AoE and GRF grants are issued by the Research Grants Council of Hong Kong S.A.R. LL is supported by the DOE grant DE-SC-0010010.

\clearpage
\bibliographystyle{jhep}
\bibliography{biblio}

\end{document}